\documentclass[aps,showpacs,nofootinbib]{revtex4}
\usepackage{latexsym}
\usepackage{amsfonts}
\usepackage{amsmath}
\usepackage{amssymb}
\usepackage{revsymb}
\usepackage{bm}
\usepackage{graphicx}
\newcommand{\bra}[1]{\langle #1 |}
\newcommand{\ket}[1]{| #1 \rangle}

\begin{document}

\begin{flushright}
VT-IPNAS-09-13
\end{flushright}

\title{Single-Coupling Bounds on R-parity violating Supersymmetry, an update}
\author{Yee Kao}\email{ykao@vt.edu}
\affiliation{Institute for Particle, Nuclear, and Astronomical Sciences,
Physics Department, Virginia Tech, Blacksburg, VA 24061}
\author{Tatsu Takeuchi}\email{takeuchi@vt.edu}
\affiliation{Institute for Particle, Nuclear, and Astronomical Sciences,
Physics Department, Virginia Tech, Blacksburg, VA 24061}

\begin{abstract}
We update the single-coupling bounds on R-parity violating supersymmetry
using the most up to date data as of October 2009.
In addition to the data listed in the 2009 Review of Particle Properties \cite{Amsler:2008zzb},
we utilize a new determination of the weak charge of cesium-133 \cite{Porsev:2009pr}, and 
preliminary $\tau$-decay branching fractions from Babar \cite{Swagato}.
Analysis of semileptonic $D$-decay is improved by the inclusion of 
experimentally measured form-factors into the calculation of the Standard Model predictions. 
\end{abstract}


\maketitle
\section{Introduction}

R-parity violating supersymmetry (SUSY) interactions \cite{RPV,Barbier:2004ez} provide a convenient framework for
quantifying quark- and lepton-flavor violating effects that new physics beyond the Standard Model (SM) may have,
independently of whether SUSY truly exists in nature or not.
Consequently, various authors have used a variety of flavor sensitive 
observables to constrain the sizes of these couplings \cite{Barbier:2004ez,nuless2beta,Huitu:1997bi,Bhattacharyya:1998be,Baek:1999ch,Lebedev:1999vc,Saha:2002kt,BarShalom:2002sv,Allanach:1999ic,Dreiner:2001kc,Kundu:2008ui,KT1}.
In this paper, we update the single-coupling bounds, namely the bounds on the individual 
R-parity violating couplings when only that particular coupling is assumed to be non-zero,
using the most up to date precision data available as of October 2009.
These include various lepton and meson decay ratios, 
CKM matrix elements, and the weak charge of atomic nuclei.

The superpotential of R-partiy violating SUSY interactions is 
given by \cite{RPV,Barbier:2004ez}
\begin{equation}
W_{\not R} =\frac{1}{2}\lambda_{ijk}\hat L_i \hat L_j \hat E_k
+\lambda^{\prime}_{ijk}\hat L_i \hat Q_j \hat D_k
+\frac{1}{2}\lambda^{\prime\prime}_{ijk}\hat U_i \hat D_j \hat D_k\;.
\end{equation}
Here $i$, $j$, $k$ are generation indices, while
$SU(2)$-weak isospin and $SU(3)$-color indices are suppressed. 
The coefficients $\lambda_{ijk}$ are antisymmetric in the first two indices, while
$\lambda''_{ijk}$ are antisymmetric in the latter two.
Consequently, there are 9 independent $LLE$ couplings, 
27 independent $LQD$ couplings, and 9 independent $UDD$ couplings.
The $\lambda''_{ijk}$ couplings lead to baryon number violating effects and
are already very strongly constrained by proton decay \cite{Barbier:2004ez}, either individually \cite{pdecay-single}
or in products with the $\lambda_{ijk}$ and $\lambda'_{ijk}$ couplings \cite{pdecay-products}, 
so they will not be considered here.

The explicit forms of the $LLE$ and $LQD$ interactions in terms of four-component spinors are
\begin{eqnarray}
\mathcal{L}_{LLE} 
& = & 
\lambda_{(i<j)k}
\biggl[\Bigl( \tilde{\nu}_{iL}\overline{e_{kR}}e_{jL}+\tilde e_{jL}\overline{e_{kR}} \nu_{iL}+
\tilde e_{kR}^{*}\overline{\nu_{iL}^c} e_{jL}\Bigr)
-
\Bigl( \tilde{\nu}_{jL}\overline{e_{kR}}e_{iL}+\tilde e_{iL}\overline{e_{kR}} \nu_{jL}+
\tilde e_{kR}^{*}\overline{\nu_{jL}^c} e_{iL}\Bigr)
\biggr]
+h.c. \cr
\mathcal{L}_{LQD} & = & \lambda_{ijk}^{\prime}
\biggl[  \left(\tilde\nu_{iL}\overline{d_{kR}} d_{jL}+
\tilde d_{jL}\overline{d_{kR}} \nu_{iL}+\tilde d_{kR}^{*}
\overline{\nu_{iL}^c} d_{jL} \right) 
-\left(
\tilde e_{iL}\overline{d_{kR}}u_{jL}+\tilde u_{jL} \overline{d_{kR}} e_{iL}+\tilde d_{kR}^* \overline{e^c_{iL}} u_{jL}
\right)
\biggr]+ h.c. 
\end{eqnarray}
Note that the charge-conjugated chiral fermion fields are denoted $f_{L/R}^c = (f_{L/R})^c = (f^c)_{R/L}$.
The exchange of squarks or sleptons mediate interactions among the SM fermions, and the 
strength of these interactions will be proportional to the ratio of coupling constant squared to the
exchanged sparticle mass squared.
To simplify or notation, we follow Ref.~\cite{Barbier:2004ez} and define
\begin{equation}
r_{ijk}(\tilde{\ell})\;\equiv\;\dfrac{1}{4\sqrt{2}G_F}\dfrac{|\lambda_{ijk}|^2}{M_{\tilde{\ell}}^2}\;,\qquad
r'_{ijk}(\tilde{q})\;\equiv\;\dfrac{1}{4\sqrt{2}G_F}\dfrac{|\lambda'_{ijk}|^2}{M_{\tilde{q}}^2}\;.
\end{equation}
Shifts in various observables will be expresses in terms of these dimensionless parameter combinations.
The final bounds on the coupling constant will be shown with all the sparticle masses set to 100~GeV.

In the following sections, we look at the bounds from
$\mu$ and $\tau$ leptonic decays, $\tau\rightarrow\pi\nu$ and $\pi$ decays, CKM unitarity, semi-leptonic $D$ and leptonic $D_s$ decays, and the weak charge of cesium-133.
For analyses involving $\tau$-decay, the impact of preliminary $\tau$-decay data from Babar \cite{Swagato} is  discussed.
The analysis of semi-leptonic $D$ decay is improved by a new calculation of the Standard Model (SM) predictions
which include the effects of experimentally determined form-factors.
The analysis of the weak charge of cesium-133 corrects an error in Ref.~\cite{Barbier:2004ez}.
Bounds from $Z$-peak observables are not updated since no new data have been generated since the 
2005 review by Barbier et al. \cite{Barbier:2004ez}.
The bound from neutrinoless double beta decay \cite{nuless2beta}
will be discussed in a separate paper \cite{KT3}.


\section{$\mu$ and $\tau$ Decay}\label{mudecay}

\begin{figure}[ht]
\begin{center}
\includegraphics{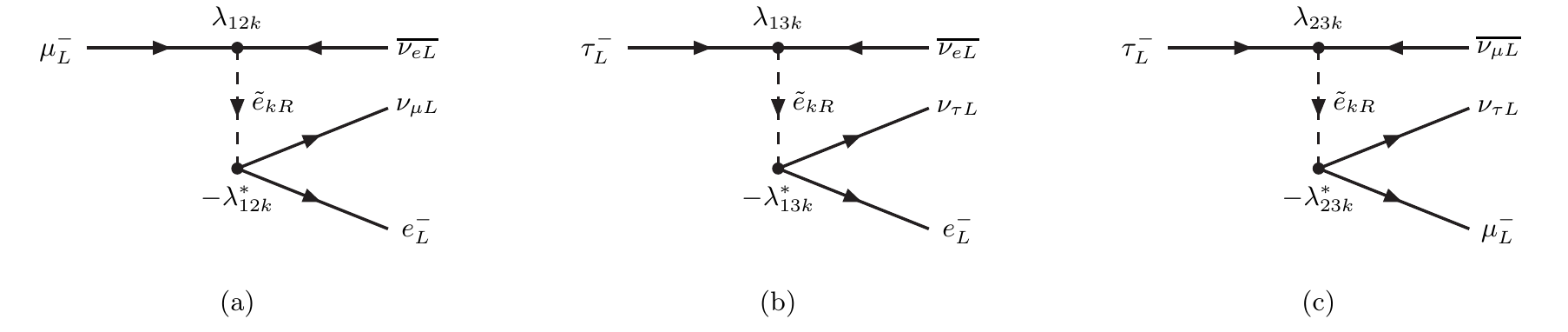}
\end{center}
\caption{Possible R-parity violating contributions to 
(a) $\mu^-\rightarrow e^-\overline{\nu_e}\nu_\mu$,
(b) $\tau^-\rightarrow e^-\overline{\nu_e}\nu_\tau$, and
(c) $\tau^-\rightarrow\mu^-\overline{\nu_\mu}\nu_\tau$.
}
\label{RPV-leptondecay}
\end{figure}

The $LLE$ couplings $\lambda_{ijk}$ affect the decays
$\mu^-\rightarrow e^-\overline{\nu_e}\nu_\mu$,
$\tau^-\rightarrow e^-\overline{\nu_e}\nu_\tau$, and
$\tau^-\rightarrow\mu^-\overline{\nu_\mu}\nu_\tau$.
via the processes shown in Fig.~\ref{RPV-leptondecay}.
The operator induced by the exchange of $\tilde{e}_{kR}$ ($k=1,2,3$) via the coupling
$\lambda_{(i<j)k}$ is 
\begin{equation}
-\dfrac{|\lambda_{(i<j)k}|^2}{M^2_{\tilde{e}_{kR}}}
\bigl(\,\overline{\nu_{iL}^c} e_{jL}^{\phantom{c}}\,\bigr)
\bigl(\,\overline{e_{iL}^{\phantom{c}}}\nu_{jL}^c\,\bigr)
\quad\stackrel{\mathrm{Fierz}}{\longrightarrow}\quad
-\dfrac{|\lambda_{(i<j)k}|^2}{2M^2_{\tilde{e}_{kR}}}
\bigl(\,\overline{\nu_{jL}} \gamma^\mu e_{jL}\,\bigr)
\bigl(\,\overline{e_{iL}}\gamma_\mu \nu_{iL}\,\bigr)\;.
\end{equation}
This will interfere with the SM operator
\begin{equation}
-\dfrac{4G_F}{\sqrt{2}}
\bigl(\,\overline{\nu_{jL}} \gamma^\mu e_{jL}\,\bigr)
\bigl(\,\overline{e_{iL}}\gamma_\mu \nu_{iL}\,\bigr)\;,\qquad
(i<j)\;,
\end{equation}
shifting the effective coupling to
\begin{equation}
\dfrac{4G_F}{\sqrt{2}} \;\rightarrow\;
\dfrac{4G_F}{\sqrt{2}} \Bigl[\, 1 + r_{(i<j)k}(\tilde{e}_{kR}) \,\Bigr]\;.
\end{equation}
In particular, $\lambda_{12k}$ will shift the muon decay constant $G_\mu$ to
\begin{equation}
G_\mu \;\rightarrow\; G_F \Bigl[\, 1 + r_{12k}(\tilde{e}_{kR}) \,\Bigr]\;,
\label{Gmushift}
\end{equation}
and this shift will also affect other observables to be discussed later.
The ratios
\begin{equation}
R_{\tau\mu}\;=\;\dfrac{\Gamma(\tau^-\rightarrow\mu^-\overline{\nu_\mu}\nu_\tau)}{\Gamma(\mu^-\rightarrow e^-\overline{\nu_e}\nu_\mu)}
\qquad\mbox{and}\qquad
R_\tau\;=\;\dfrac{\Gamma(\tau^-\rightarrow e^-\overline{\nu_e}\nu_\tau)}{\Gamma(\tau^-\rightarrow \mu^-\overline{\nu_\mu}\nu_\tau)}
\end{equation}
will be shifted to
\begin{eqnarray}
R_{\tau\mu} & = & [R_{\tau\mu}]_{\mathrm{SM}}
\biggl[ 1 + 2
\Bigl\{ 
 r_{23k}(\tilde{e}_{kR})
-r_{12k}(\tilde{e}_{kR})
\Bigr\}
\biggr] \;, \cr
R_{\tau} & = & [R_{\tau}]_{\mathrm{SM}}\;
\biggl[ 1 + 2
\Bigl\{
 r_{13k}(\tilde{e}_{kR})
-r_{23k}(\tilde{e}_{kR})
\Bigr\}
\biggr] \;.
\label{RtauRPV}
\end{eqnarray}
The SM predictions for the decay widths
including radiative corrections \cite{Kinoshita:1958ru,Marciano:vm}
are:
\begin{eqnarray}
\Gamma(\tau\rightarrow \mu\, \bar{\nu}_\mu\, \nu_\tau\,(\gamma) )_{\mathrm{SM}}
& = & \frac{g^4}{64 M_W^4}\,
      \frac{ m_\tau^5 }{ 96\pi^3}\;
       f\!\left(\frac{m_\mu^2}{m_\tau^2}\right)
      \delta^\tau_W\, \delta^\tau_\gamma\;, \cr
\Gamma(\tau\rightarrow e\, \bar{\nu}_e\, \nu_\tau\,(\gamma) )_{\mathrm{SM}}
& = & \frac{g^4}{64 M_W^4}\,
      \frac{ m_\tau^5 }{ 96\pi^3}\;
       f\!\left(\frac{m_e^2}{m_\tau^2}\right)
      \delta^\tau_W\, \delta^\tau_\gamma\;, \cr
\Gamma(\mu\rightarrow e\, \bar{\nu}_e\, \nu_\mu\,(\gamma) )_{\mathrm{SM}}
& = & \frac{g^4}{64 M_W^4}\,
      \frac{ m_\mu^5 }{ 96\pi^3}\;
       f\!\left(\frac{m_e^2}{m_\mu^2}\right)
      \delta^\mu_W\, \delta^\mu_\gamma
\;=\; \frac{1}{\tau_\mu}\;, 
\end{eqnarray}
in which $f(x)$ is the phase space factor
\begin{equation}
f(x) = 1 - 8x + 8x^3 - x^4 - 12 x^2\ln x\;,
\end{equation}
$\delta^\ell_W$ is the $W$ propagator correction
\begin{equation}
\delta^\ell_W = \left( 1 +\frac{3}{5}\frac{m_\ell^2}{M_W^2}
                \right),
\end{equation}
$\delta^\ell_\gamma$ is the radiative correction from photons
\begin{equation}
\delta^\ell_\gamma = 
1 + \frac{\alpha(m_\ell)}{2\pi}\left( \frac{25}{4} - \pi^2 \right)\;,
\end{equation}
and the values of the running QED coupling constant at the
relevant energies are \cite{Marciano:vm}
\begin{eqnarray}
\alpha^{-1}(m_\mu) & = & \alpha^{-1} 
-\frac{2}{3\pi}\ln\frac{m_\mu}{m_e} 
+\frac{1}{6\pi}
\approx 136.0\;, \cr
\alpha^{-1}(m_\tau) & \approx & 133.3 \;.
\end{eqnarray}
The numerical values of these corrections are shown in Table~\ref{taumucorrections}.
\begin{table}
\begin{tabular}{l||c|c|c}
& phase space & $W$ propagator & photon \\
\hline
$\Gamma(\tau\rightarrow \mu\,\bar{\nu}_\mu\, \nu_\tau)$ &
$\;f(m_\mu^2/m_\tau^2) = 0.9726\;$ & 
$\;\delta^\tau_W = 1.0003\;$ & 
$\;\delta^\tau_\gamma = 0.9957\;$ \\
\cline{1-2}
$\Gamma(\tau\rightarrow e\,\bar{\nu}_e\, \nu_\tau)$ &
$\;f(m_e^2/m_\tau^2) = 1.0000\;$ & & \\
\hline
$\Gamma(\mu\rightarrow e\,\bar{\nu}_e\, \nu_\mu)$ &
$\;f(m_e^2/m_\mu^2) = 0.9998\;$ & 
$\;\delta^\mu_W = 1.0000\;$ & 
$\delta^\mu_\gamma = 0.9958$ \\
\hline
\end{tabular}
\caption{The corrections to the leptonic decay widths of the $\tau$ and $\mu$.}
\label{taumucorrections}
\end{table}
The SM predictions for the ratios are therefore
\begin{eqnarray}
[R_{\tau\mu}]_{\mathrm{SM}} & = &
\dfrac{\Gamma(\tau\rightarrow\mu\,\bar{\nu}_\mu\,\nu_\tau\,(\gamma))_{\mathrm{SM}}}
      {\Gamma(\mu\rightarrow e\,\bar{\nu}_e\,\nu_\mu\,(\gamma))_{\mathrm{SM}}}
\;=\; \frac{m_\tau^5}{m_\mu^5}\;
      \frac{f(m_\mu^2/m_\tau^2)}
           {f(m_e^2  /m_\mu^2)}\;
      \frac{\delta^\tau_W}{\delta^\mu_W}\;
      \frac{\delta^\tau_\gamma}{\delta^\mu_\gamma}
\;=\; 1.309\times 10^6\;,\cr
[R_\tau]_{\mathrm{SM}} & = &
\dfrac{\Gamma(\tau\rightarrow e\,\bar{\nu}_e\,\nu_\tau\,(\gamma))_{\mathrm{SM}}}
      {\Gamma(\tau\rightarrow\mu\,\bar{\nu}_\mu\,\nu_\tau\,(\gamma))_{\mathrm{SM}}}
\;=\; \frac{f(m_e^2    /m_\tau^2)}
           {f(m_\mu^2  /m_\tau^2)}
\;=\; 1.028\;.
\label{RtauSM}
\end{eqnarray}
The experimental values of these ratios from the Review of Particle Properties \cite{Amsler:2008zzb} are
\begin{eqnarray}
R_{\tau\mu} & = &
\dfrac{\tau_\mu}{\tau_\tau}\,
\mathcal{B}(\tau\rightarrow\mu\,\bar{\nu}_\mu\,\nu_\tau\,(\gamma))
\;=\; \dfrac{(2.197034\pm 0.000021)\times 10^{-6}\,\mathrm{s}}{(290.6\pm 1.0)\times 10^{-15}\,\mathrm{s}}\;
(17.36\pm 0.05)\%
\;=\; (1.312\pm 0.006)\times 10^6\;,\cr
R_{\tau} & = &
\dfrac{\mathcal{B}(\tau\rightarrow e\,\bar{\nu}_e\,\nu_\tau\,(\gamma))}
      {\mathcal{B}(\tau\rightarrow\mu\,\bar{\nu}_\mu\,\nu_\tau\,(\gamma))}
\;=\; 
\dfrac{(17.85\pm 0.05)\%}{(17.36\pm 0.05)\%}
\;=\; 1.028\pm 0.004\;.
\label{RtauExp}
\end{eqnarray}
The effect of a $-13\%$ correlation between $\mathcal{B}(\tau\rightarrow e\,\bar{\nu}_e\,\nu_\tau\,(\gamma))$ and
$\mathcal{B}(\tau\rightarrow\mu\,\bar{\nu}_\mu\,\nu_\tau\,(\gamma))$ on the error on $R_\tau$ is small.
Allowing only one of the $\lambda$'s to be non-zero at a time, 
comparison of Eqs.~(\ref{RtauRPV}), (\ref{RtauSM}), and (\ref{RtauExp}) places the following $2\sigma$ bounds:
\begin{eqnarray}
|\lambda_{12k}|\left(\dfrac{100\,\mathrm{GeV}}{M_{\tilde{e}_{kR}}}\right) & < & 0.05\;[R_{\tau\mu}]\;,\cr
|\lambda_{13k}|\left(\dfrac{100\,\mathrm{GeV}}{M_{\tilde{e}_{kR}}}\right) & < & 0.05\;[R_\tau]\;,\cr
|\lambda_{23k}|\left(\dfrac{100\,\mathrm{GeV}}{M_{\tilde{e}_{kR}}}\right) & < & 0.05\;[R_\tau]\;,\;0.06\;[R_{\tau\mu}]\;.
\end{eqnarray}
A new but still preliminary value of $R_\tau$ from Babar was announced at ICHEP 2008 \cite{Swagato} as
\begin{equation}
[R_\tau]_{\mathrm{Babar2008}} \;=\; \dfrac{1}{0.9796\pm 0.0038} \;=\; 1.021\pm 0.004\;.
\label{BabarNewRtau}
\end{equation}
Including this value will change the world average to
\[
R_\tau \;=\; 1.025\pm 0.003\;,
\]
and the corresponding $2\sigma$ bounds will be
\begin{equation}
|\lambda_{13k}|\left(\dfrac{100\,\mathrm{GeV}}{M_{\tilde{e}_{kR}}}\right) \;<\; 0.03\;[R_\tau]\;,\qquad
|\lambda_{23k}|\left(\dfrac{100\,\mathrm{GeV}}{M_{\tilde{e}_{kR}}}\right) \;<\; 0.05\;[R_\tau]\;.
\end{equation}
We see that the bound on $\lambda_{13k}$ will be tightened.

\section{$\pi$ and $\tau$ Decay}

\begin{figure}[ht]
\begin{center}
\includegraphics{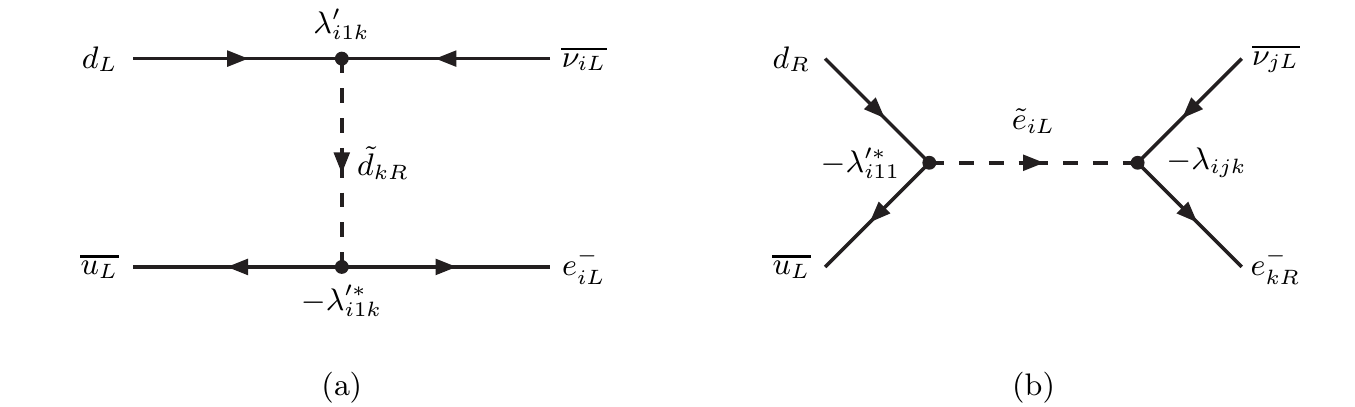}
\end{center}
\caption{Possible R-parity violating contributions to 
$\pi^-\rightarrow \ell^- \overline{{\nu}_\ell}$ 
($\ell=e_1=e$, or $e_2=\mu$) that interfere with the SM amplitude.
The indices are
$i=1$ or $2$,
$k=1$, $2$, or $3$ in (a); while $(jk)=(11)$ or $(22)$, with $i=3-j$ or $3$ in (b) due to the anti-symmetry of $\lambda_{ijk}$
in the first two indices.
The interference of (b) with the SM amplitude is suppressed due to the smallness of the electron mass.}
\label{RPV-pidecay}
\end{figure}

\begin{figure}[ht]
\begin{center}
\includegraphics{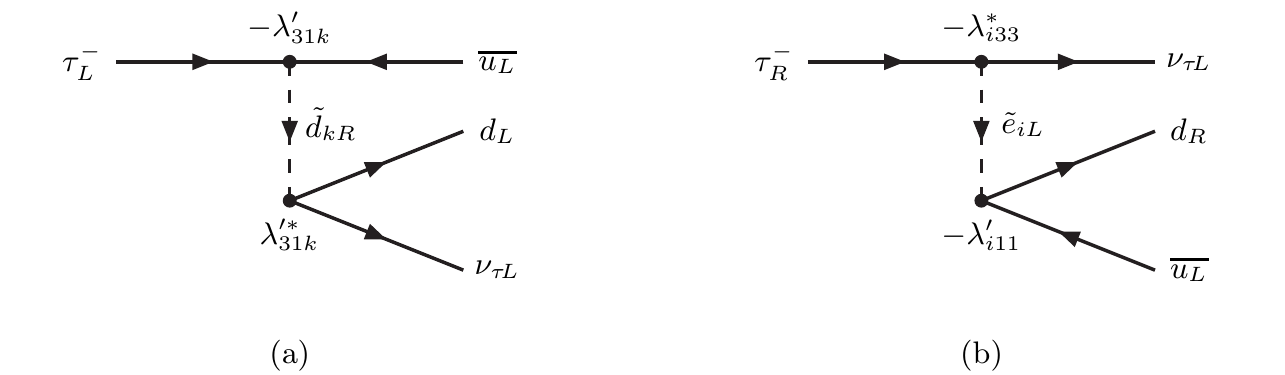}
\end{center}
\caption{Possible R-parity violating contributions to 
the decay $\tau^-\rightarrow \pi^- \nu_{\tau}$.
}
\label{RPV-tau2pi}
\end{figure}

Possible R-parity violating contributions to the decay $\pi^-\rightarrow \ell^-\overline{{\nu}_\ell}$ 
($\ell=e$ or $\mu$)
are shown in Fig.~\ref{RPV-pidecay}, and those to $\tau^-\rightarrow \pi^-\nu_\tau$ are shown in 
Fig.~\ref{RPV-tau2pi}.  Since we are only interested in placing bounds on the individual R-parity violating
couplings separately, we will ignore the (b) diagrams in both cases.

The processes of Fig.~\ref{RPV-pidecay}(a) and Fig.~\ref{RPV-tau2pi}(a) induce the following operators:
\begin{eqnarray}
\ref{RPV-pidecay}(a) : & \quad & 
-\dfrac{|\lambda'_{i1k}|^2}{M^2_{\tilde{d}_{kR}}}
\bigl(\,\overline{\nu_{iL}^c} d_{jL}^{\phantom{c}}\,\bigr)
\bigl(\,\overline{u_{L}^{\phantom{c}}}e_{iL}^c\,\bigr)
\quad\stackrel{\mathrm{Fierz}}{\longrightarrow}\quad
-\dfrac{|\lambda_{i1k}|^2}{2M^2_{\tilde{d}_{kR}}}
\bigl(\,\overline{\nu_{iL}^c} \gamma^\mu e_{iL}^c\,\bigr)
\bigl(\,\overline{u_{L}}\gamma_\mu d_{L}\,\bigr)
\;=\;
-\dfrac{|\lambda_{i1k}|^2}{2M^2_{\tilde{d}_{kR}}}
\bigl(\,\overline{e_{iL}} \gamma^\mu \nu_{iL}\,\bigr)
\bigl(\,\overline{u_{L}}\gamma_\mu d_{L}\,\bigr)
\;,
\cr
\ref{RPV-tau2pi}(a) : & \quad &
-\dfrac{|\lambda'_{31k}|^2}{M^2_{\tilde{d}_{kR}}}
\bigl(\,\overline{\tau_{L}^c} u_{L}^{\phantom{c}}\,\bigr)
\bigl(\,\overline{d_{L}}\nu_{\tau L}^c\,\bigr)
\quad\stackrel{\mathrm{Fierz}}{\longrightarrow}\quad
-\dfrac{|\lambda_{31k}|^2}{2M^2_{\tilde{d}_{kR}}}
\bigl(\,\overline{\tau_{L}^c} \gamma^\mu \nu_{\tau L}^c\,\bigr)
\bigl(\,\overline{d_{L}}\gamma_\mu u_{L}\,\bigr)
\;=\;
-\dfrac{|\lambda_{31k}|^2}{2M^2_{\tilde{d}_{kR}}}
\bigl(\,\overline{\nu_{\tau L}} \gamma^\mu \tau_{L}\,\bigr)
\bigl(\,\overline{d_{L}}\gamma_\mu u_{L}\,\bigr)
\;.
\cr
& &
\end{eqnarray}
These interfere with the SM operators given by
\begin{equation}
-\dfrac{4G_F}{\sqrt{2}}V_{ud}
\bigl(\,\overline{e_{iL}} \gamma^\mu \nu_{iL}\,\bigr)
\bigl(\,\overline{u_{L}}\gamma_\mu d_{L}\,\bigr)
\;,\qquad\mbox{and}\qquad
-\dfrac{4G_F}{\sqrt{2}}V_{ud}^*
\bigl(\,\overline{\nu_{\tau L}} \gamma^\mu \tau_{L}\,\bigr)
\bigl(\,\overline{d_{L}}\gamma_\mu u_{L}\,\bigr)
\;,
\end{equation}
and shift the $\pi$-decay widths to
\begin{equation}
\Gamma(\pi^-\rightarrow \ell_i^- \overline{{\nu}_{\ell_i}})
\;=\; 
\left[\,\Gamma(\pi^-\rightarrow \ell_i^- \overline{{\nu}_{\ell_i}})\,\right]_{\mathrm{SM}}
\left[\; 1 + \dfrac{2}{|V_{ud}|}\,r'_{i1k}(\tilde{d}_{kR})
\;\right]\;,
\end{equation}
while the $\tau$-decay width is shifted by
\begin{equation}
\Gamma(\tau^-\rightarrow\pi^-\nu_\tau)
\;=\;
\left[\,
\Gamma(\tau^-\rightarrow\pi^-\nu_\tau)
\,\right]_{\mathrm{SM}}
\left[\,
1 + \dfrac{2}{|V_{ud}|}\, r'_{31k}(\tilde{d}_{kR})
\,\right]\;.
\end{equation}
Here, we have neglected any relative phase between the SM and RPV contributions.
The ratios
\begin{eqnarray}
R_\pi & = &  \dfrac{\Gamma(\pi^-\rightarrow e^-\overline{\nu_e})}{\Gamma(\pi^-\rightarrow\mu^-\overline{\nu_\mu})}
\;,\cr
R_{\tau\pi} & = & \dfrac{\Gamma(\tau^-\rightarrow \pi^-\nu_\tau)}{\Gamma(\pi^-\rightarrow\mu^-\overline{\nu_\mu})}
\;,
\end{eqnarray}
are shifted to
\begin{eqnarray}
R_\pi & = & [R_\pi]_{\mathrm{SM}}
\left[\; 1 + \dfrac{2}{|V_{ud}|}
\left\{r'_{11k}(\tilde{d}_{kR})-r'_{21k}(\tilde{d}_{kR})
\right\}
\right]\;,\cr
R_{\tau\pi} & = & 
[\,R_{\tau\pi}\,]_\mathrm{SM}
\left[\,
1 + \dfrac{2}{|V_{ud}|}
\left\{\,
  r'_{21k}(\tilde{d}_{kR})
- r'_{31k}(\tilde{d}_{kR})
\,\right\}
\,\right]\;.
\label{RpiRPV}
\end{eqnarray}
At tree level, the SM prediction for the $\pi$-decay widths is given by
\begin{equation}
[\,\Gamma(\pi^-\rightarrow \ell^-\,\overline{{\nu}_\ell})\,]_{\mathrm{SM,tree}}
\;=\; \left(\sqrt{2}G_F |V_{ud}| \right)^2 \dfrac{m_\ell^2 m_\pi}{16\pi}
\left(1-\frac{m_\ell^2}{m_\pi^2}\right)^2 f_\pi^2\;, 
\end{equation}
while that of $\tau$-decay into $\pi\,\nu_\tau$ is
\begin{equation}
[\,\Gamma(\tau^-\rightarrow \pi^-\,\nu_\tau)\,]_{\mathrm{SM,tree}}
\;=\; \left(\sqrt{2}G_F |V_{ud}| \right)^2 \dfrac{m_\tau^3}{32\pi}
\left(1-\frac{m_\pi^2}{m_\tau^2}\right)^2 f_\pi^2 \;,
\end{equation}
where the pion decay constant $f_\pi$ is normalized as 
\begin{equation}
\langle 0 |\,\bar{u}\gamma_\mu \gamma_5 d \,(0)\,| \pi^-(\mathbf{q}) \rangle = i\,q_\mu f_\pi\;.
\end{equation}
Taking ratios, we find
\begin{eqnarray}
[\,R_{\pi}\,]_{\mathrm{SM,tree}}
& = &
\dfrac{[\,\Gamma(\pi^-\rightarrow e^- \,\overline{\nu_e}  )\,]_{\mathrm{SM,tree}}}
      {[\,\Gamma(\pi^-\rightarrow\mu^-\,\overline{\nu_\mu})\,]_{\mathrm{SM,tree}}}
\;=\;
\frac{m_e^2}{m_\mu^2}\,
\frac{(1-m_e^2  /m_\pi^2)^2}
     {(1-m_\mu^2/m_\pi^2)^2} 
\;=\; 1.283\times 10^{-4}    
\;, \cr
[\,R_{\tau\pi}\,]_{\mathrm{SM,tree}} 
& = &
\frac{[\,\Gamma(\tau^-\rightarrow\pi^-\,\nu_\tau     )\,]_{\mathrm{SM,tree}}}
     {[\,\Gamma(\pi^- \rightarrow\mu^-\,\overline{\nu_\mu})\,]_{\mathrm{SM,tree}}}
\;=\; 
\frac{m_\tau^3}{2 m_\mu^2 m_\pi}\;
\frac{(1-m_\pi^2/m_\tau^2)^2}
     {(1-m_\mu^2 /m_\pi^2)^2}
\;=\; 9.756\times 10^3
\;.
\end{eqnarray}
Radiative corrections to these relations have been calculated
in Ref.~\cite{Decker:1994ea} and modify them to
\begin{eqnarray}
[\,R_\pi\,]_{\mathrm{SM}} & = & 
[\,R_\pi\,]_{\mathrm{SM,tree}}
\left(1+\delta R_{\pi}\right) \;, \cr
[\,R_{\tau\pi}\,]_{\mathrm{SM}} & = &
[\,R_{\tau\pi}\,]_{\mathrm{SM,tree}}
\left(1+\delta R_{\tau\pi}\right) \;,
\end{eqnarray}
with
\begin{equation}
\delta R_{\pi} = -0.0374\pm 0.0001 \;,\qquad
\delta R_{\tau\pi} = +0.0016^{+0.0009}_{-0.0014}\;.
\end{equation}
The uncertainty in these corrections is due to
the uncertainty from strong interaction effects.
Therefore,
\begin{eqnarray}
[\,R_\pi\,]_{\mathrm{SM}} & = & 1.235\times 10^{-4}\;,\cr
[\,R_{\tau\pi}\,]_{\mathrm{SM}} & = & 9.771^{+0.009}_{-0.013}\times 10^3\;.
\label{RpitauSM}
\end{eqnarray}
On the other hand,
the current experimental values are \cite{Amsler:2008zzb}
\begin{eqnarray}
R_\pi & = & 
\dfrac{\mathcal{B}(\pi\rightarrow e \,\bar{\nu}_e  \,(\gamma))}
     {\mathcal{B}(\pi\rightarrow\mu\,\bar{\nu}_\mu\,(\gamma))}
\;=\; \dfrac{(0.01230\pm 0.00004)\%}{(99.98770\pm 0.00004)\%}
\;=\; (1.230\pm 0.004) \times 10^{-4}\;,
\cr
R_{\tau\pi} & = &
\frac{\tau_\pi}{\tau_\tau}\,
\frac{\mathcal{B}(\tau\rightarrow\pi \,\nu_\tau     \,(\gamma))}
     {\mathcal{B}(\pi \rightarrow\mu\,\bar{\nu}_\mu \,(\gamma))}
\;=\;
\dfrac{(2.6033\pm 0.0005)\times 10^{-8}\mathrm{s}}{(290.6\pm 1.0)\times 10^{-15}\mathrm{s}}
\;\dfrac{(10.91\pm 0.07)\%}{(99.98770\pm 0.00004)\%}
\;=\; (9.775\pm 0.071)\times 10^3\;.
\cr & &
\label{RpitauExp}
\end{eqnarray}
The magnitude of the CKM matrix element $V_{ud}$ is \cite{Ceccucci:2008zz},
\begin{equation}
|V_{ud}| \;=\; 0.97418 \pm 0.00027\;.
\label{Vudvalue}
\end{equation}
Comparison of Eqs.~(\ref{RpiRPV}), (\ref{RpitauSM}) and (\ref{RpitauExp}) leads to the following $2\sigma$ bounds
assuming only one of the couplings is non-zero at a time:
\begin{eqnarray}
|\lambda'_{11k}| \left(\dfrac{100\,\mathrm{GeV}}{M_{\tilde{d}_{kR}}}\right) & < & 0.03\;[R_\pi]\;,\cr 
|\lambda'_{21k}| \left(\dfrac{100\,\mathrm{GeV}}{M_{\tilde{d}_{kR}}}\right) & < & 0.06\;[R_\pi]\;,\; 0.07\;[R_{\tau\pi}]\;,\cr 
|\lambda'_{31k}| \left(\dfrac{100\,\mathrm{GeV}}{M_{\tilde{d}_{kR}}}\right) & < & 0.06\;[R_{\tau\pi}]\;.
\end{eqnarray}
Another preliminary result announced at ICHEP 2008 from Babar \cite{Swagato} was
\begin{equation}
\left[\dfrac{\mathcal{B}(\tau^-\rightarrow \pi^-\nu_\tau)}{\mathcal{B}(\tau^-\rightarrow e^-\bar{\nu}_e\nu_\tau)}
\right]_{\mathrm{Babar2008}}
\;=\; 0.5945 \pm 0.0063\;.
\end{equation}
Using the current world average value of
$\mathcal{B}(\tau^-\rightarrow e^-\bar{\nu}_e\nu_\tau)=(17.85\pm 0.05)\%$
\cite{Amsler:2008zzb} we find
\begin{equation}
\bigl[\,\mathcal{B}(\tau^-\rightarrow \pi^-\nu_\tau)\,\bigr]_{\mathrm{Babar2008}} \;=\;  10.61\pm 0.12\;.
\end{equation}
Including this value will shift the world average to
\begin{equation}
\mathcal{B}(\tau^-\rightarrow \pi^-\nu_\tau) \;=\;  10.83\pm 0.06\;,
\end{equation}
and the ratio $R_{\tau\pi}$ to
\begin{equation}
R_{\tau\pi} \;=\; (9.703\pm 0.063)\times 10^3\;.
\end{equation}
The error will be reduced somewhat and the central value shifted down by about $1\sigma$. 
The $2\sigma$ bounds will become
\begin{equation}
|\lambda'_{21k}| \left(\dfrac{100\,\mathrm{GeV}}{M_{\tilde{d}_{kR}}}\right)\;<\; 0.04\;[R_{\tau\pi}]\;,\qquad
|\lambda'_{31k}| \left(\dfrac{100\,\mathrm{GeV}}{M_{\tilde{d}_{kR}}}\right)\;<\; 0.08\;[R_{\tau\pi}]\;,
\end{equation}
the change mostly due to the shift in the central value of $R_{\tau\pi}$.

\section{CKM unitarity}

\begin{figure}[ht]
\begin{center}
\includegraphics{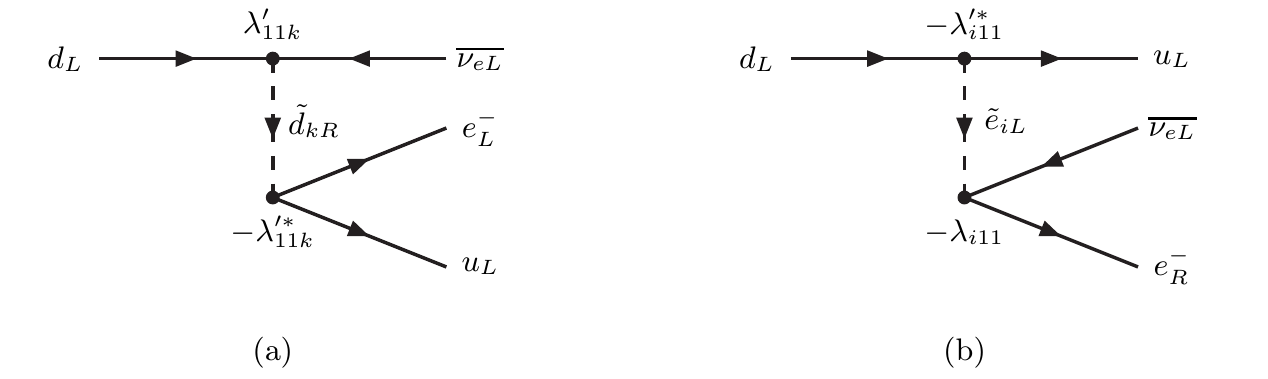}
\end{center}
\caption{Possible R-parity violating contributions to 
nuclear beta decay.}
\label{RPV-betadecay}
\end{figure}

\begin{figure}[ht]
\begin{center}
\includegraphics{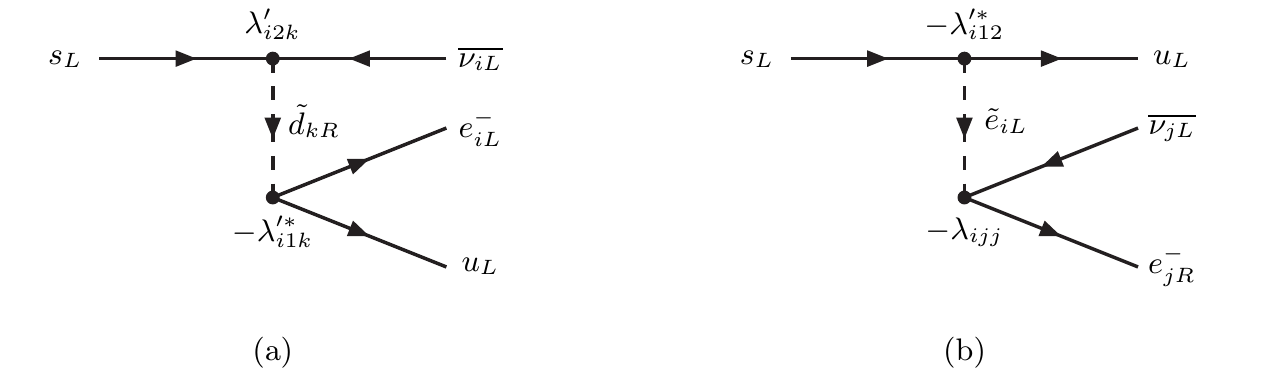}
\end{center}
\caption{Possible R-parity violating contributions to 
semileptonic $K$-decay.
}
\label{RPV-Kdecay}
\end{figure}

The SM values of the CKM matrix elements $V_{ud}$, $V_{us}$, and $V_{ub}$ must satisfy the
unitarity relation
\begin{equation}
|V_{ud}^{\mathrm{SM}}|^2 + 
|V_{us}^{\mathrm{SM}}|^2 +
|V_{ub}^{\mathrm{SM}}|^2
\;=\; 1\;.
\end{equation}
Deviation of the measured values from this relation could be a sign of new physics.
The value of $|V_{ud}|$, cited above in Eq.~(\ref{Vudvalue}), is obtained from the comparison of 
superallowed $0^+\rightarrow 0^+$ nuclear beta decays and muon decay \cite{Ceccucci:2008zz},
the former used to extract the product $G_F V_{ud}$ and the latter used to cancel the $G_F$.
The SM operators relevant for these decays are
\begin{equation}
-\dfrac{4G_F}{\sqrt{2}} V_{ud}^{\mathrm{SM}}
\bigl(\overline{u_L}\gamma^\mu d_L\bigr)
\bigl(\overline{e_L}\gamma_\mu \nu_{eL}\bigr)\;,
\qquad\mbox{and}\qquad
-\dfrac{4G_F}{\sqrt{2}}
\bigl(\overline{\nu_{\mu L}}\gamma^\mu \mu_L\bigr)
\bigl(\overline{e_L}\gamma_\mu \nu_{eL}\bigr)\;.
\end{equation} 
Any new physics amplitude which interferes with these operators will affect the extracted value of $V_{ud}$.
The first operator is the same as the operator responsible for the decay $\pi^-\rightarrow e^-\overline{\nu_e}$.
The RPV amplitudes which interfere with this are the same as
those shown in Fig.~\ref{RPV-pidecay} except with the $u_L$ lines pointing toward the future
as shown in Fig.~\ref{RPV-betadecay}.  Since Fig.~\ref{RPV-betadecay}(b) depends on two different
RPV couplings (its interference is also suppressed by the electron mass)
we will only consider Fig.~\ref{RPV-betadecay}(a) for which
the corresponding operator is
\begin{equation}
-\dfrac{|\lambda'_{11k}|^2}{M_{\tilde{d}_{kR}}^2}
\bigl(\overline{u_L^{\phantom{c}}}e_L^c\bigr)
\bigl(\overline{\nu_{eL}^c}d_L^{\phantom{c}}\bigr)
\quad\stackrel{\mathrm{Fierz}}{\longrightarrow}\quad
+\dfrac{|\lambda'_{11k}|^2}{2M_{\tilde{d}_{kR}}^2}
\bigl(\overline{u_L^{\phantom{c}}}\gamma^\mu d_L^{\phantom{c}}\bigr)
\bigl(\overline{\nu_{eL}^c}\gamma_\mu e_L^c\bigr)
\;=\; 
-\dfrac{|\lambda'_{11k}|^2}{2M_{\tilde{d}_{kR}}^2}
\bigl(\overline{u_L^{\phantom{c}}}\gamma^\mu d_L^{\phantom{c}}\bigr)
\bigl(\overline{e_L}\gamma_\mu \nu_{eL}\bigr)\;.
\end{equation}
We have already discussed muon decay in section~\ref{mudecay} where we found that
$G_F$ will be shifted by $r_{12k}(\tilde{e}_{kR})$, \textit{cf.} Eq.~(\ref{Gmushift}).
%
Therefore, the shift in $V_{ud}$ will be
\begin{equation}
|V_{ud}|^2 \;=\; 
|V_{ud}^\mathrm{SM}|^2
\left[
1 
+ \dfrac{2}{|V_{ud|}}\,r'_{11k}(\tilde{d}_{kR})
- 2\,r_{12k}(\tilde{e}_{kR})
\right] \;.
\end{equation}
The values of $V_{us}$ and $V_{ub}$ extracted from semi-leptonic $K$ and $B$ decays are
\cite{Ceccucci:2008zz,Kowalewski:2008zz}
\begin{eqnarray}
|V_{us}| & = & 0.2255 \pm 0.0019 \;, \cr
|V_{ub}| & = & (3.95\pm 0.35)\times 10^{-3}\;.
\end{eqnarray}
The RPV diagrams that contribute to semi-leptonic $K$-decay are shown in
Fig.~\ref{RPV-Kdecay}.  Similar diagrams contribute to semi-leptonic $B$-decay.
None of these diagrams depend on a single RPV coupling so we may neglect them
and assume
\begin{equation}
V_{us}^{\phantom{S}} \;=\; V_{us}^\mathrm{SM}\;,\qquad
V_{ub}^{\phantom{S}} \;=\; V_{ub}^\mathrm{SM}\;.
\end{equation}
Then,
\begin{equation}
\left[
1 
+ \dfrac{2}{|V_{ud}|}\,r'_{11k}(\tilde{d}_{kR})
- 2\,r_{12k}(\tilde{e}_{kR})
\right]
\;=\;
\dfrac{|V_{ud}|^2}{1-|V_{us}|^2-|V_{ub}|^2}
\;=\; 0.9999\pm 0.0011\;.
\end{equation}
The $2\sigma$ bounds on the couplings are
\begin{equation}
|\lambda'_{11k}|\left(\dfrac{100\,\mathrm{GeV}}{M_{\tilde{d}_{kR}}}\right)
\;<\; 0.03\;[V_{ud}]\;,\qquad
|\lambda_{12k}|\left(\dfrac{100\,\mathrm{GeV}}{M_{\tilde{e}_{kR}}}\right)
\;<\; 0.03\;[V_{ud}]\;.
\end{equation}
%

\section{Semi-leptonic $D$ and leptonic $D_s$-decay}

\begin{figure}[ht]
\begin{center}
\includegraphics{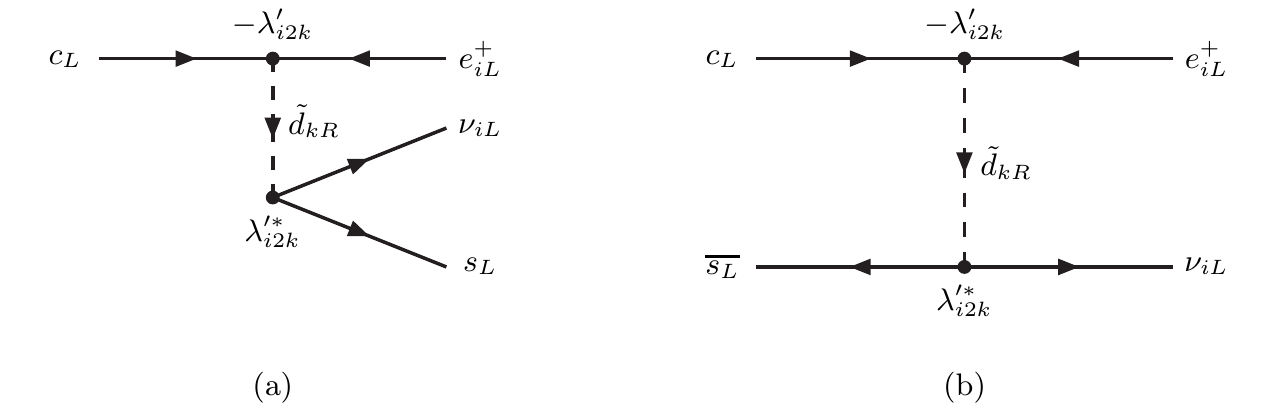}
\end{center}
\caption{Possible R-parity violating contributions to 
semileptonic $D$-decay and leptonic $D_s$-decay.
}
\label{RPV-Ddecay}
\end{figure}

The process shown in Fig.~\ref{RPV-Ddecay}(a) affects semileptonic $D$-decay,
while that in Fig.~\ref{RPV-Ddecay}(b) affects leptonic $D_s$-decay.
They are both described by the same operator given by
\begin{equation}
-\dfrac{|\lambda'_{i2k}|^2}{M^2_{\tilde{d}_{kR}}}
\bigl(\,\overline{e_{iL}^c} c_{L}^{\phantom{c}}\,\bigr)
\bigl(\,\overline{s_{L}}\nu_{i L}^c\,\bigr)
\quad\stackrel{\mathrm{Fierz}}{\longrightarrow}\quad
-\dfrac{|\lambda_{i2k}|^2}{2M^2_{\tilde{d}_{kR}}}
\bigl(\,\overline{e_{iL}^c} \gamma^\mu \nu_{i L}^c\,\bigr)
\bigl(\,\overline{s_{L}}\gamma_\mu c_{L}\,\bigr)
\;=\;
-\dfrac{|\lambda_{i2k}|^2}{2M^2_{\tilde{d}_{kR}}}
\bigl(\,\overline{\nu_{i L}} \gamma^\mu e_{iL}\,\bigr)
\bigl(\,\overline{s_{L}}\gamma_\mu c_{L}\,\bigr)
\;.
\end{equation}
This interferes with the SM operator
\begin{equation}
-\dfrac{4G_F}{\sqrt{2}}V_{cs}^*
\bigl(\,\overline{\nu_{i L}} \gamma^\mu e_{iL}\,\bigr)
\bigl(\,\overline{s_{L}}\gamma_\mu c_{L}\,\bigr)
\;,
\end{equation}
shifting the $D$ and $D_s$ decay widths by
\begin{equation}
\dfrac{\Gamma(D\rightarrow K\ell_i\nu_{\ell_i})}
{\bigl[\,\Gamma(D\rightarrow K\ell_i\nu_{\ell_i})\,\bigr]_{\mathrm{SM}}}
\;=\;
\dfrac{\Gamma(D\rightarrow K^*\ell_i\nu_{\ell_i})}
{\bigl[\,\Gamma(D\rightarrow K^*\ell_i\nu_{\ell_i})\,\bigr]_{\mathrm{SM}}}
\;=\;
\dfrac{\Gamma(D_s\rightarrow \ell_i\nu_{\ell_i})}
{\bigl[\,\Gamma(D_s\rightarrow \ell_i\nu_{\ell_i})\,\bigr]_{\mathrm{SM}}}
\;=\;
1 + \dfrac{2}{|V_{cs}|}\, r'_{i2k}(\tilde{d}_{kR})
\;,
\end{equation}
where we have neglected any relative phase between the SM and RPV contributions.
Following Ref.~\cite{Barbier:2004ez}, we define the ratios
\begin{eqnarray}
R_{D^0} & = &
\dfrac{\mathcal{B}(D^0\rightarrow \mu^+ \nu_\mu K^-)}
      {\mathcal{B}(D^0\rightarrow e^+ \nu_e K^-)}\;,\cr
R_{D^+} & = &
\dfrac{\mathcal{B}(D^+\rightarrow \mu^+ \nu_\mu \overline{K^0})}
      {\mathcal{B}(D^+\rightarrow e^+ \nu_e \overline{K^0})}\;,\cr
R_{D^+}^* & = &
\dfrac{\mathcal{B}(D^+\rightarrow \mu^+ \nu_\mu \overline{K^*}(892)^0)}
      {\mathcal{B}(D^+\rightarrow e^+ \nu_e \overline{K^*}(892)^0)}\;,
\label{RDdefs}
\end{eqnarray}
the shifts of which are
\begin{equation}
\dfrac{R_{D^0}}{[R_{D^0}]_{\mathrm{SM}}} \;=\;
\dfrac{R_{D^+}}{[R_{D^+}]_{\mathrm{SM}}} \;=\;
\dfrac{R_{D^+}^*}{[R_{D^+}^*]_{\mathrm{SM}}} \;=\;
1 + 
\dfrac{2}{|V_{cs}|}
\left\{
 r'_{22k}(\tilde{d}_{kR})
-r'_{12k}(\tilde{d}_{kR})
\right\}\;.
\label{RDshift}
\end{equation}
The experimental values of these ratios are currently \cite{Amsler:2008zzb}
\begin{eqnarray}
R_{D^0} & = & 
\dfrac{(3.32\pm 0.13)\times 10^{-2}}{(3.61\pm 0.05)\times 10^{-2}}
\;=\; 0.92 \pm 0.04\;,\cr
R_{D^+} & = & 
\dfrac{(9.4\pm 0.8)\times 10^{-2}}{(8.50\pm 0.26)\times 10^{-2}}
\;=\; 1.1\pm 0.1\;,\cr
R_{D^+}^* & = & 
\dfrac{(5.4\pm 0.5)\times 10^{-2}}{(5.51\pm 0.31)\times 10^{-2}}
\;=\; 0.98\pm 0.10\;.
\label{RDexp}
\end{eqnarray}
Calculating the SM predictions of these ratios requires knowledge of the form-factors for
the matrix elements \cite{DdecayFormFactors}
\begin{equation}
\bra{K}\,\overline{s}\gamma^\mu(1-\gamma_5)c\,\ket{D}\qquad\mbox{and}\qquad
\bra{K^*}\,\overline{s}\gamma^\mu(1-\gamma_5)c\,\ket{D}\;,
\end{equation}
for which good experimental data now exist from 
FOCUS \cite{Link:2004dh,Link:2002wg,Link:2004uk}, 
Belle \cite{Widhalm:2006wz}, 
Babar \cite{Aubert:2007wg}, and CLEO \cite{Dobbs:2007sm,Besson:2009uv}.
Details of our calculation are presented in the appendix.
The results are:
\begin{eqnarray}
\bigl[\,R_{D^0}\,\bigr]_\mathrm{SM}^{-1}
\;=\;
\bigl[\,R_{D^+}\,\bigr]_\mathrm{SM}^{-1}
& = & 1.04\pm 0.02\;(1\sigma)\;,\quad 1.04^{+0.02}_{-0.06}\;(2\sigma)\;,\cr
\bigl[\,R^*_{D^+}\,\bigr]_\mathrm{SM}^{-1}
& = & 1.060^{+0.002}_{-0.003}\;(1\sigma)\;,\quad 1.060^{+0.005}_{-0.007}\;(2\sigma)\;.
\label{RDSM}
\end{eqnarray}
The analysis of Ref.~\cite{Barbier:2004ez} used the value of $(1.03)^{-1}$ without any errors for all three ratios.
This would be the value of $R_{D^0}$ and $R_{D^+}$ if form factors are ignored and the $D$ and the $K$ treated as point particles with the interaction $(K^\dagger\!\stackrel{\leftrightarrow}{\partial_\mu}\!\!D)W^\mu$, and it lies within our calculated range above.
For the ratio $R^*_{D^+}$, if form factors are ignored and the $D$ and the $K^*$ treated as point particles
with the interaction $K^{*\dagger}_\mu D\,W^\mu$, its value would be $(1.12)^{-1}$, 
which illustrates the importance of taking form-factors into account.
The value of $|V_{cs}|$, also extracted from $D$ semileptonic decays and $D_s$ leptonic decays
\cite{Ceccucci:2008zz}, is 
\begin{equation}
|V_{cs}| \;=\; 1.04\pm 0.06\;,
\end{equation}
and for our current purpose we can set it to one.
In comparing Eqs.~(\ref{RDshift}), (\ref{RDexp}), and (\ref{RDSM}),  
we allow the SM predictions to scan the entire $2\sigma$ range of Eq.~(\ref{RDSM}) 
and pick up the weakest bounds on the couplings.
The resulting $2\sigma$ bounds are
\begin{eqnarray}
|\lambda'_{12k}|\left(\dfrac{100\,\mathrm{GeV}}{M_{\tilde{d}_{kR}}}\right) 
& < & 0.2\;[R_{D^0}]\;,\;0.2\;[R_{D^+}]\;,\;\; 0.2\;[R_{D^+}^*]\;,\cr
|\lambda'_{22k}|\left(\dfrac{100\,\mathrm{GeV}}{M_{\tilde{d}_{kR}}}\right) 
& < & 0.1\;[R_{D^0}]\;,\;0.4\;[R_{D^+}]\;,\;\; 0.3\;[R_{D^+}^*]\;.
\end{eqnarray}

\bigskip

Next, define
\begin{equation}
R_{D_s}(\tau\mu) \;=\;
\dfrac{\mathcal{B}(D_s^+ \rightarrow \tau^+ \nu_\tau)}
      {\mathcal{B}(D_s^+ \rightarrow \mu^+ \nu_\mu)}\;.
\end{equation}
The shift of this ratio is
\begin{equation}
\dfrac{R_{D_s}(\tau\mu)}{\bigl[\,R_{D_s}(\tau\mu)\,\bigr]_{\mathrm{SM}}}
\;=\; 
1 + \dfrac{2}{|V_{cs}|}
\left\{r'_{32k}(\tilde{d}_{kR})
      -r'_{22k}(\tilde{d}_{kR})
\right\}
\;.
\end{equation}
The current experimental value is
\begin{equation}
R_{D_s}(\tau\mu) \;=\;
\dfrac{(6.6\pm 0.5)\times 10^{-2}}{(6.3\pm 0.5)\times 10^{-3}}
\;=\; 10.5\pm 1.1\;,
\end{equation}
while the tree-level SM prediction is
\begin{equation}
R_{D_s}(\tau\mu) \;=\;
\dfrac{m_\tau^2}{m_\mu^2}
\dfrac{(1-m_\tau^2/m_{D_s}^2)^2}{(1-m_\mu^2/m_{D_s}^2)^2}
\;=\; 9.76 \pm 0.03\;.
\end{equation}
Comparison of the two leads to the
$2\sigma$ bounds given by
\begin{equation}
|\lambda'_{22k}|\left(\dfrac{100\,\mathrm{GeV}}{M_{\tilde{d}_{kR}}}\right) \;<\; 0.2\;[R_{D_s}(\tau\mu)]\;,\qquad
|\lambda'_{32k}|\left(\dfrac{100\,\mathrm{GeV}}{M_{\tilde{d}_{kR}}}\right) \;<\; 0.3\;[R_{D_s}(\tau\mu)]\;.
\end{equation}

\section{Atomic Parity Violation}

\begin{figure}[ht]
\begin{center}
\includegraphics{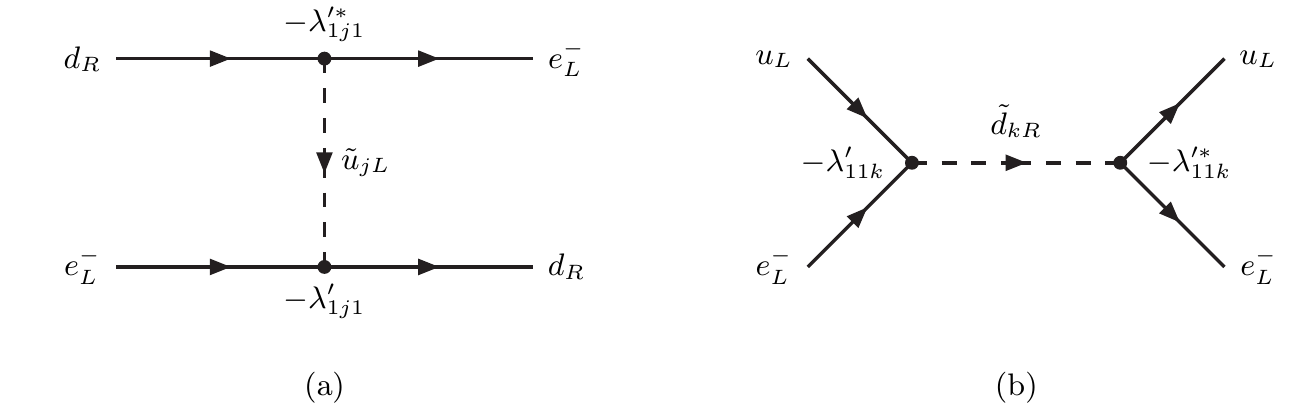}
\end{center}
\caption{Possible R-parity violating contributions to 
atomic parity violation.}
\label{APV}
\end{figure}

The diagrams shown in Fig.~(\ref{APV}) lead to the following effective couplings between the
quarks and the electron:
\begin{eqnarray}
(a) : &\quad& \dfrac{|\lambda'_{11k}|^2}{M_{\tilde{d}_{kR}}^2}
\bigl(\overline{e_{L}^c}u_L\bigr)
\bigl(\overline{u_L}e_L^c\bigr)
\quad\stackrel{\mathrm{Fierz}}{\longrightarrow}\quad
-
\dfrac{|\lambda'_{11k}|^2}{2M_{\tilde{d}_{kR}}^2}
\bigl(\overline{e_{L}^c}\gamma_\mu e_L^c\bigr)
\bigl(\overline{u_L}\gamma^\mu u_L\bigr)
\;=\; 
\dfrac{|\lambda'_{11k}|^2}{2M_{\tilde{d}_{kR}}^2}
\bigl(\overline{e_{L}}\gamma_\mu e_L\bigr)
\bigl(\overline{u_L}\gamma^\mu u_L\bigr)
\;,\cr
(b) : &\quad& \dfrac{|\lambda'_{1j1}|^2}{M_{\tilde{u}_{jL}}^2}
\bigl(\overline{e_L}d_R\bigr)
\bigl(\overline{d_R}e_L\bigr)
\quad\stackrel{\mathrm{Fierz}}{\longrightarrow}\quad
-
\dfrac{|\lambda'_{1j1}|^2}{2M_{\tilde{u}_{jL}}^2}
\bigl(\overline{e_L}\gamma^\mu e_L\bigr)
\bigl(\overline{d_R}\gamma_\mu d_R\bigr)
\;.
\end{eqnarray}
The parity violating parts of these interactions are
\begin{eqnarray}
(a):\quad & & -
\dfrac{|\lambda'_{11k}|^2}{8M_{\tilde{d}_{kR}}^2}
\Bigl[
 \bigl(\overline{e}\gamma_\mu \gamma_5 e\bigr)\bigl(\overline{u}\gamma^\mu u\bigr)
+\bigl(\overline{e}\gamma_\mu e\bigr)\bigl(\overline{u}\gamma^\mu \gamma_5 u\bigr)
\Bigr]
\;,\cr
(b):\quad & & \phantom{-}
\dfrac{|\lambda'_{1j1}|^2}{8M_{\tilde{u}_{jL}}^2}
\Bigl[
 \bigl(\overline{e}\gamma_\mu \gamma_5 e\bigr)\bigl(\overline{d}\gamma^\mu d\bigr)
-\bigl(\overline{e}\gamma_\mu e\bigr)\bigl(\overline{d}\gamma^\mu \gamma_5 d\bigr)
\Bigr]
\;.
\end{eqnarray}
The parity violating SM interactions, on the other hand, are
\begin{equation}
\dfrac{G_F}{\sqrt{2}}\sum_{q=u,d}
\Bigl[
 C_{1q}\bigl(\overline{e}\gamma_\mu \gamma_5 e\bigr)\bigl(\overline{q}\gamma^\mu q\bigr)
+C_{2q}\bigl(\overline{e}\gamma_\mu e\bigr)\bigl(\overline{q}\gamma^\mu \gamma_5 q\bigr)
\Bigr]\;,
\end{equation}
with
\begin{eqnarray}
C_{1u} & = & -\dfrac{1}{2}+\dfrac{4}{3}s^2\;,\qquad
C_{2u} \;=\; -\dfrac{1}{2}+2s^2 \;,
\cr
C_{1d} & = & +\dfrac{1}{2}-\dfrac{2}{3}s^2\;,\qquad
C_{2d} \;=\; +\dfrac{1}{2}-2s^2 \;,
\end{eqnarray}
at tree-level,
$s^2$ being the shorthand for $\sin^2\theta_W$.
The weak charge of an atomic nucleus with $Z$ protons and $N$ neutrons is given by
\begin{equation}
Q_W(Z,N)
\;=\; -2
\Bigl[ C_{1u}(2Z+N)+C_{1d}(Z+2N) \Bigr]
\;=\; (1-4s^2)Z - N \;.
\end{equation}
Note that $(2Z+N)$ and $(Z+2N)$ are respectively the number of up and down quarks in the nucleus.
Since the presence of the above R-parity violating couplings will shift the $C_1$ couplings to
\begin{eqnarray}
C_{1u} & \rightarrow & C_{1u} - r'_{11k}(\tilde{d}_{kR})\;,\cr
C_{1d} & \rightarrow & C_{1d} + r'_{1j1}(\tilde{u}_{jL})\;,
\end{eqnarray}
the weak charge will be shifted by
\begin{equation}
\delta Q_W(Z,N) \;=\;
+2\Big[ (2Z+N)\,r'_{11k}(\tilde{d}_{kR})-(Z+2N)\,r'_{1j1}(\tilde{u}_{jL})\Bigr]\;.
\end{equation}
Furthermore, since the quantity that is actually measured in atomic parity violation (APV)
experiments is the product $G_F Q_W$, and $Q_W$ is extracted by setting $G_F$ equal to the muon decay constant
$G_\mu$, the $LLE$ coupling $r_{12k}(\tilde{e}_{kR})$ can also affect $Q_W$ via Eq.~(\ref{Gmushift}):
\begin{equation}
\dfrac{G_F Q_W}{G_\mu}
\;=\; Q_W\bigl[\,1-r_{12k}(\tilde{e}_{kR})\,\bigr]\;.
\end{equation}
For cesium-133, the total shift will be
\begin{equation}
\delta Q_W(55,78)
\;=\; 376\,r'_{11k}(\tilde{d}_{kR})-422\,r'_{1j1}(\tilde{u}_{jL})
- Q_W(55,78)\,r_{12k}(\tilde{e}_{kR})\;.
\end{equation}
(This formula differs from that provided on page 82 of Ref.~\cite{Barbier:2004ez} 
which contains a typo: the factor of $-2$ on the right-hand-side should not be there.
This error seems to have propagated into the bounds listed in Eq.~(6.47) on the same page.
We correct for this error in quoting the bounds from Ref.~\cite{Barbier:2004ez} in Table~\ref{Bounds}.)
The latest experimental value of the weak charge of cesium-133 is \cite{Porsev:2009pr}
\begin{equation}
Q_W({}^{133}_{\phantom{0}55}\mathrm{Cs}) \;=\; -73.16\pm 0.29 \pm 0.20 \;=\; -73.16\pm 0.35\;.
\end{equation}
On the other hand,
the SM value provided in the Review of Particle Properties \cite{Amsler:2008zzb} is 
\begin{equation}
\bigl[\,Q_W({}^{133}_{\phantom{0}55}\mathrm{Cs})\,\bigr]_\mathrm{SM}
\;=\; -73.16\pm 0.03\;,
\end{equation}
which is based on a global fit to all electroweak observables
with radiative corrections only from within the SM.
Assuming further radiative corrections from new physics is negligibly small, 
and saturating the difference between the two with the RPV contributions, we find
the following $2\sigma$ bounds:
\begin{equation}
\lambda'_{11k}\left(\dfrac{100\,\mathrm{GeV}}{M_{\tilde{d}_{kR}}}\right)
\;<\;0.04\;[Q_W]\;,\qquad
\lambda'_{1j1}\left(\dfrac{100\,\mathrm{GeV}}{M_{\tilde{u}_{jL}}}\right)
\;<\;0.03\;[Q_W]\;,\qquad
\lambda_{12k}\left(\dfrac{100\,\mathrm{GeV}}{M_{\tilde{e}_{kR}}}\right)
\;<\;0.08\;[Q_W]\;.
\end{equation}
%

\section{Summary}

\begin{table}[ht]
\begin{tabular}{|c||c||l||l|l|}
\hline
\ Coupling\ \ 
& \ Observable\ \
& \ Previous $2\sigma$ bound \cite{Barbier:2004ez}\ \
& \ New $2\sigma$ bound \ \ 
& \ with Babar preliminary \cite{Swagato}\ \ \\
\hline\hline
$\lambda_{12k}$ & $V_{ud}$           & \ $0.05$   & \ $0.03$ & \\
                & $R_{\tau\mu}$      & \ $0.07$   & \ $0.05$ & \\
                & $Q_W(\mathrm{Cs})$ & \ $0.11^*$ & \ $0.08$ & \\
\hline
$\lambda_{13k}$ & $R_\tau$           & \ $0.07$   & \ $0.05$ & \ $0.03$ \\
\hline
$\lambda_{23k}$ & $R_\tau$           & \ $0.07$   & \ $0.05$ & \ $0.05$ \\
                & $R_{\tau\mu}$      & \ $0.07$   & \ $0.06$ & \ \\ 
\hline\hline
$\lambda'_{1j1}$ & $Q_W(\mathrm{Cs})$ & \ $0.03^*$ & \ $0.03$ & \\
\hline
$\lambda'_{11k}$ & $V_{ud}$           & \ $0.02$   & \ $0.03$ & \\
                 & $R_\pi$            & \ $0.03$   & \ $0.03$ & \\
                 & $Q_W(\mathrm{Cs})$ & \ $0.05^*$ & \ $0.04$ & \\
\hline
$\lambda'_{12k}$ & $R_{D^0}$          & \ $0.27^\dagger$ & \ $0.2$ & \\
                 & $R_{D^+}$          & \ $0.44^\dagger$ & \ $0.2$ & \\
                 & $R^*_{D^+}$        & \ $0.23^\dagger$ & \ $0.2$ & \\
\hline
$\lambda'_{21k}$ & $R_\pi$            & \ $0.06$   & \ $0.06$ & \ \\
                 & $R_{\tau\pi}$      & \ $0.08$   & \ $0.07$ & \ $0.04$ \\
\hline
$\lambda'_{22k}$ & $R_{D^0}$          & \ $0.21^\dagger$ & \ $0.1$ & \\
                 & $R_{D^+}$          & \ $0.61^\dagger$ & \ $0.4$ & \\
                 & $R^*_{D^+}$        & \ $0.38^\dagger$ & \ $0.3$ & \\
                 & $R_{D_s}(\tau\mu)$ & \ $0.65$   & \ $0.2$ & \\
\hline
$\lambda'_{31k}$ & $R_{\tau\pi}$      & \ $0.12$   & \ $0.06$ & \ $0.08$ \\
\hline
$\lambda'_{32k}$ & $R_{D_s}(\tau\mu)$ & \ $0.52$   & \ $0.3$ & \\
\hline
\end{tabular}
\caption{Current single-coupling bounds on various RPV couplings compared with those given in
Table~6.1 of Ref.~\cite{Barbier:2004ez}.
All sparticle masses have been set to 100~GeV.
The values with an asterisk have been corrected for an error in Ref.~\cite{Barbier:2004ez}.
Those with a dagger were calculated with the SM prediction fixed to $(1.03)^{-1}$ without any uncertainties.
} 
\label{Bounds}
\end{table}

In this paper,
we have looked at a variety of single-coupling bounds that can be imposed on R-parity violation from
particle decay ratios and atomic parity violation.
Our results are summarized in Table~\ref{Bounds}.
Compared to the bounds complied in 2004 by Barbier et al. \cite{Barbier:2004ez}, the bounds have steadily improved.
For the bounds from observables involving $\tau$-decay,
the improvement is due to the results of ALEPH \cite{Schael:2005am} released in 2005.
The bounds from observables involving $D$ and $D_s$ decay have improved due to new data from
CLEO \cite{Dobbs:2007sm,Artuso:2007zg,Pedlar:2007za,Ecklund:2007zm} and Belle \cite{Widhalm:2006wz,Widhalm:2007ws}.
The new bounds from the weak charge of cesium-133 is due to the reduction of theoretical uncertainty in 
its extraction from experimental data \cite{Porsev:2009pr}.

Further improvements on the bounds from observables involving $\tau$-decay is expected
as analyses of Belle and Babar data get under way, each with hundreds of millions of $\tau^+\tau^-$ pairs, as is evidenced by the effect of the preliminary Babar data from 2008 \cite{Swagato}.
The bounds from observables involving $\pi$-decay can also be expected to improve considerably in the near future
as the PIENU experiment at TRIUMF \cite{PIENU} and the PEN experiment and PSI \cite{Pocanic:2009gd}
start their physics runs this year, eventually improving the error on $R_\pi$ by almost an order of magnitude.

The bound on $\lambda'_{111}$ from neutrinoless double beta decay will be updated in a subsequent paper \cite{KT3}.

\section*{Acknowledgements}

This work was supported by the U.S. Department of Energy, 
grant DE--FG05--92ER40709, Task A.

\appendix
\section{Semileptonic $D$ decay ratios with form-factors}

The semileptonic decays $D\rightarrow K\ell\nu$ ($D^+\rightarrow \overline{K^0}\ell^+\nu_\ell$ or $D^0\rightarrow K^-\ell^+\nu_\ell$)
and $D\rightarrow K^*\ell\nu$ ($D^+\rightarrow\overline{K}^*(892)^0\ell^+\nu_\ell$ or $D^0\rightarrow K^*(892)^-\ell^+\nu_\ell$)
proceed via the SM interaction
\begin{equation}
\mathcal{L} \;=\; -\dfrac{G_F}{\sqrt{2}}V_{cs}^*
\bigl[\,\bar{s}\gamma_\mu(1-\gamma_5)c\,\bigr]
\bigl[\,\bar{\nu}\gamma^\mu(1-\gamma_5)\ell\,\bigr]\;.
\end{equation}
For both cases, the hadronic matrix element is expressed in terms of multiple form factors.
In this appendix, we calculate the ratios $R_{D^0}$, $R_{D^+}$, and $R^*_{D^+}$,
defined in Eq.~(\ref{RDdefs}), taking the latest experimental 
data on these form factors into account.
%

\subsection{$D\rightarrow K\ell\nu$}

For the decay $D\rightarrow K\ell\nu$, the form factors are defined as \cite{DdecayFormFactors}
\begin{equation}
\bra{K}\bar{s}\gamma_\mu(1-\gamma_5)c\ket{D}
\;=\; 
(P_D + P_K)_\mu\, f_+(t) +
(P_D - P_K)_\mu\, f_-(t)\;,
\end{equation}
where $t=(P_D-P_K)^2$.  The decay width in terms of these form factors is
%
%
%
\begin{eqnarray}
\Gamma
& = & \dfrac{G_F^2|V_{cs}|^2 m_D^3}{24\pi^3}
\int_{m_\ell^2}^{(m_D-m_K)^2} dt\;
\Biggl[\,
\left\{\dfrac{\lambda^{1/2}(t,m_D^2,m_K^2)}{2m_D^2}\right\}^3
\left(1-\dfrac{m_\ell^2}{t}\right)^2 \left(1+\dfrac{m_\ell^2}{2t}\right) |f_+(t)|^2 
\cr
& & \hspace{4.5cm}
+\dfrac{3}{8}\,\dfrac{m_\ell^2}{m_D^2}
\left\{\dfrac{\lambda^{1/2}(t,m_D^2,m_K^2)}{2m_D^2}\right\}
\left(1-\dfrac{m_\ell^2}{t}\right)^2 
\dfrac{t}{m_D^2}\left(\dfrac{m_D^2-m_K^2}{t}\right)^2
|f_0(t)|^2
\Biggr]
\;,
\label{GammaDtoK}
\end{eqnarray}
where we have defined
\begin{equation}
f_0(t) \;\equiv\; f_+(t) + \dfrac{t}{m_D^2-m_K^2}\;f_-(t)\;,
\end{equation}
and
\begin{equation}
\lambda(a,b,c) \;=\; a^2 + b^2 + c^2 - 2ab - 2bc - 2ca\;.
\end{equation}
Since $m_e^2\ll m_D^2$, the scalar form factor term is
completely negligible for the electron, and
the vector form factor $|f_+(t)|$ can be extracted from the $t$-dependence of $d\Gamma/dt$.
The result is fit by the single-pole function
\begin{equation}
f_+(t)\;=\;\dfrac{f_+(0)}{1-t/m_\mathrm{pole}^2}\;,
\end{equation}
where $f_+(0)$ and $m_\mathrm{pole}$ are adjustable parameters.
This form is motivated by the vector meson dominance model \cite{Sakurai:1969ss} which 
prescribes the following expressions for the form factors \cite{Finkemeier:1996dh}:
\begin{eqnarray}
f_+(t) & = & c_V\,\dfrac{1}{1-t/m_V^2}\;,\cr
f_0(t) & = & c_V + c_S\, \dfrac{t/m_S^2}{1-t/m_S^2}\;.
\label{VMD}
\end{eqnarray}
Here, $c_V$ and $c_S$ are constants, and $m_V$ and $m_S$ are the masses of the
lowest lying $D_s$-mesons with $J^P=1^-$ and $0^+$, respectively.
Another popular form used to fit the $f_+(t)$ data is the so-called modified-pole function 
\begin{equation}
f_+(t)\;=\;\dfrac{f_+(0)}{(1-t/m_{D_s^*}^2)(1-\alpha\,t/m_{D_s^*}^2)}\;,
\end{equation}
where $m_{D_s^*}=2112\,\mathrm{MeV}$, and $f_+(0)$ and $\alpha$ are adjustable.
The fit values of $m_\mathrm{pole}$ and $\alpha$ from recent experiments are listed in
Table~\ref{DtoKformfactors}.

\begin{table}[ht]
\begin{tabular}{|c||l|l|l|l|l|}
\hline
\ Collaboration\ \ &\ \ Decay Mode\ \ &\ \ $m_\mathrm{pole}$ (GeV)\ \ &\ \ $\alpha$\ \ &\ \ $f_-(0)/f_+(0)$\ \ &\ Reference\ \  \\
\hline
FOCUS
&\ $D^0\rightarrow K^- \mu^+\nu_\mu$\ \
&\ $1.93\pm 0.05\pm 0.03$\ \
&\ $0.28\pm 0.08\pm 0.07$\ \
&\ $-1.7^{+1.5}_{-1.4}\pm 0.3$\ \
&\ \cite{Link:2004dh}\ 2005\ \\
\hline
Belle
&\ $D^0\rightarrow \overline{K}^- \ell^+\nu_\ell$ ($\ell=e,\mu$)\ \
&\ $1.82\pm 0.04\pm 0.02$\ \
&\ $0.52\pm 0.08\pm 0.06$\ \
&
&\ \cite{Widhalm:2006wz}\ 2006\ \\
\hline
Babar
&\ $D^0\rightarrow K^- e^+\nu_e$
&\ $1.884\pm 0.012\pm 0.016$\ \
&\ $0.377\pm 0.023\pm 0.031$\ \
&
&\ \cite{Aubert:2007wg}\ 2007\ \\
\hline
CLEO
&\ $D^0\rightarrow K^- e^+\nu_e$, $D^+\rightarrow \overline{K}^0 e^+\nu_e$\ \ 
&\ $1.93\pm 0.02\pm 0.01$\ \ 
&\ $0.30\pm 0.03\pm 0.01$\ \ 
&
&\ \cite{Besson:2009uv}\ 2009\ \\
\hline
\end{tabular}
\caption{Recent measurements of the $D\rightarrow K$ form factors.}
\label{DtoKformfactors}
\end{table}

\begin{figure}[ht]
\includegraphics[width=8cm]{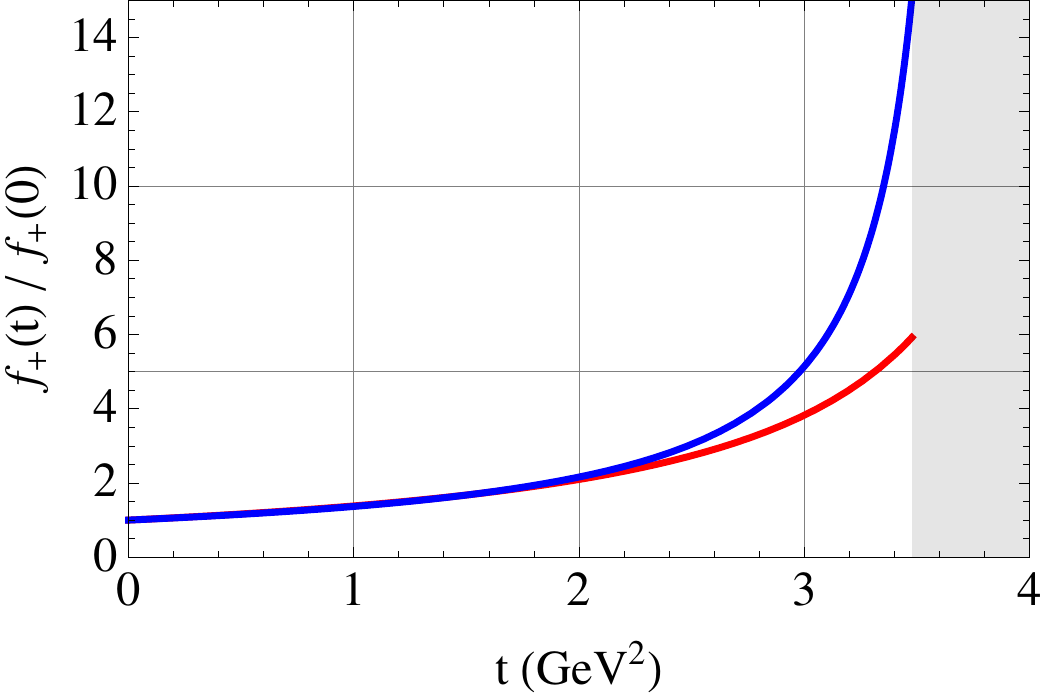}
\caption{The single-pole function with $m_\mathrm{pole}=1.93$~GeV (solid), and
the modified-pole function with $\alpha=0.30$ (dashed).
The end-point is at $t=m_{D}^2\approx 3.5\,\mathrm{GeV}^2$.}
\label{ModPole}
\end{figure}

The forms of the single-pole and modified-pole functions for the CLEO \cite{Besson:2009uv}
central values of $m_\mathrm{pole}$ and $\alpha$ are shown in Fig.~\ref{ModPole}.
They deviate from each other considerably beyond $t \sim 2\,\mathrm{GeV}^2$, where data points are absent due to phase-space
suppression.  However, due to this same phase-space suppression, it turns out that the $R_{D}$ ratios are 
insensitive to this difference in the shape of $f_+(t)$ in this region of $t$,
so we adopt the single-pole form for our purpose. 

\begin{figure}[h]
\includegraphics[width=8cm]{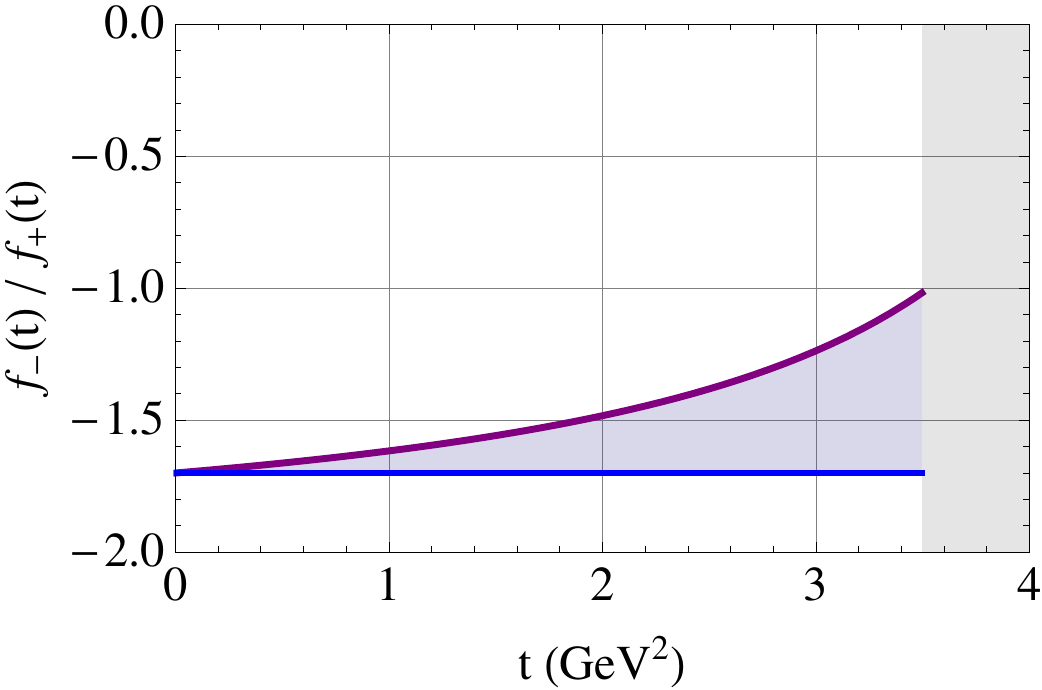}
\caption{The ratio $f_-(t)/f_+(t)$ with $m_S = m_\mathrm{pole}=1930\,\mathrm{MeV}$ (solid), 
and $m_S = m(D^*_{s0}(2317)^\pm)=2317\,\mathrm{MeV}$ (dashed).
The FOCUS central value is used for the value of ratio at $t=0$.}
\label{fratio}
\end{figure}

FOCUS \cite{Link:2004dh}, which looked at the muon channel, has
determined the ratio $f_-(0)/f_+(0)$, in addition to $m_\mathrm{pole}$ and $\alpha$,  
by performing a two-dimensional fit to 
$d^2\Gamma/dt\, d\!\cos\theta_\ell$, 
where $\theta_\ell$ is the angle between the
neutrino and the kaon in the $\mu\nu$ rest-frame.
If we adopt the meson-dominance form given in Eq.~(\ref{VMD}) for $f_0(t)$, 
with $c_V=f_+(0)$, 
this ratio determines $r_0 = c_S/c_V$ as
\begin{equation}
r_0 \;=\; \dfrac{c_S}{c_V} \;=\;
\dfrac{m_S^2}{m_\mathrm{pole}^2}+\dfrac{m_S^2}{m_D^2-m_K^2}\,\dfrac{f_-(0)}{f_+(0)}\;.
\end{equation}
In the FOCUS analysis, it was assumed that the ratio $f_-(t)/f_+(t)$ 
was essentially independent of $t$, which would require $m_S\approx m_\mathrm{pole}$.
However, as noted above, data points do not extend into the regions of $t$ which are phase-space suppressed,
thus $m_S$ can be differ significantly from $m_\mathrm{pole}$ while still maintaining an almost constant
ratio $f_-(t)/f_+(t)$ in the regions measured.  
Indeed, in Fig.~\ref{fratio} we show how this ratio varies with $t$ for the two choices 
$m_S = m_\mathrm{pole}=1930\,\mathrm{MeV}$ and 
and $m_S = m(D^*_{s0}(2317)^\pm)=2317\,\mathrm{MeV}$ when $f_-(0)/f_+(0)=-1.7$.
In the region $t<2\,\mathrm{GeV}^2$, the ratio only varies from $-1.7$ to about $-1.5$ 
even for the latter case.
Thus, in our calculations, we allow $m_S$ to vary between $m_\mathrm{pole}$ and $m(D^*_{s0}(2317)^\pm)$.

\begin{figure}[h]
\includegraphics[width=9cm]{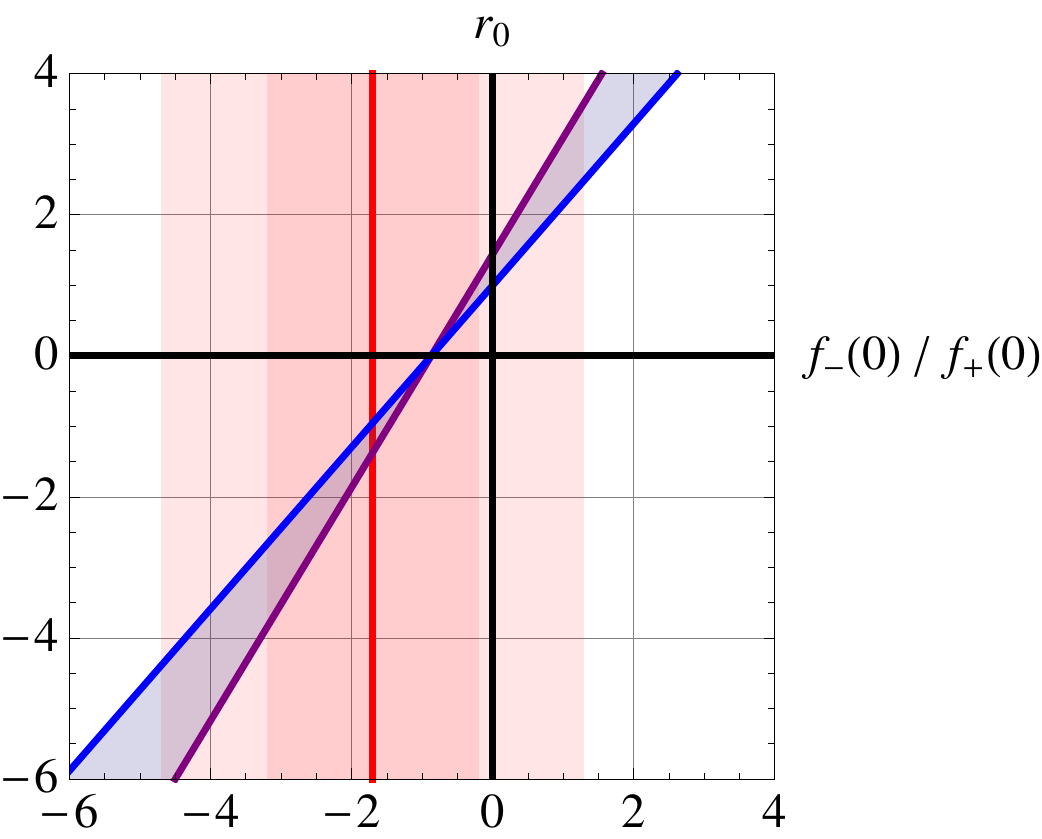}
\caption{The dependence of $r_0$ on the ratio $f_-(0)/f_+(0)$
with $m_S = m_\mathrm{pole}=1930\,\mathrm{MeV}$ (solid), 
and $m_S = m(D^*_{s0}(2317)^\pm)=2317\,\mathrm{MeV}$ (dashed).
The shaded bands indicate the one- and two-sigma regions of $f_-(0)/f_+(0)$ as measured by FOCUS \cite{Link:2004dh}.}
\label{rzero}
\end{figure}

The form-factors we use are therefore,
\begin{eqnarray}
\dfrac{f_+(t)}{f_+(0)} & = & \dfrac{1}{1-t/m_\mathrm{pole}^2} \;,\cr
\dfrac{f_0(t)}{f_+(0)} & = & 1 + r_0\,\dfrac{t/m_S^2}{1-t/m_S^2}\;.
\end{eqnarray}
For $m_\mathrm{pole}$ we adopt the FOCUS value of $1.93\,\mathrm{GeV}$, which coincides with the
CLEO value.  The errors on $m_\mathrm{pole}$ are small enough as to have no effect on $R_D$ since they only change the form of $f_+(t)$ near the pole. The value of $r_0$ will depend on the value of the ratio $f_-(0)/f_+(0)$
and our choice of $m_S$ as shown in Fig.~\ref{rzero}.

\begin{figure}[h]
\includegraphics[width=8cm]{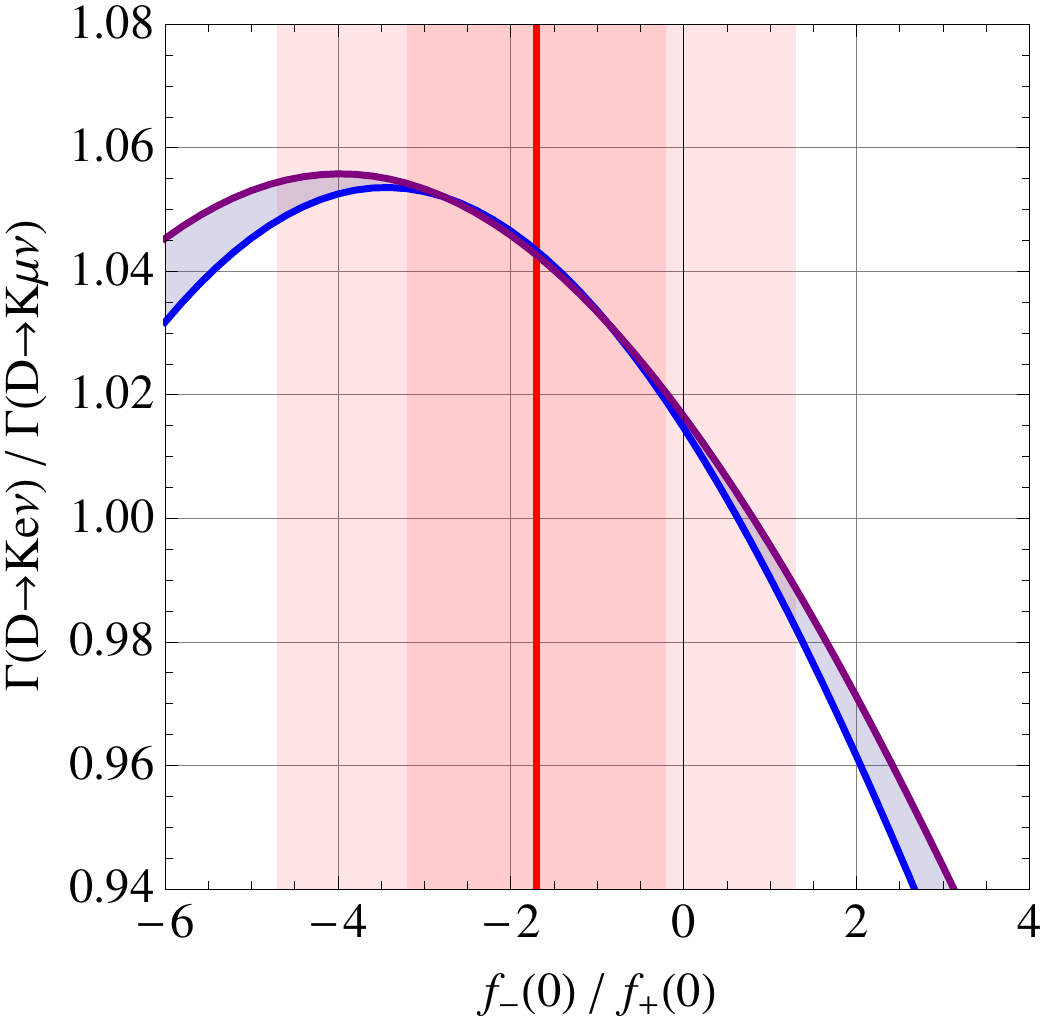}
\caption{The dependence of $(R_{D})^{-1}=\Gamma(D\rightarrow Ke\nu)/\Gamma(D\rightarrow K\mu\nu)$ 
on the ratio $f_-(0)/f_+(0)$
with $m_S = m_\mathrm{pole}=1930\,\mathrm{MeV}$ (solid), 
and $m_S = m(D^*_{s0}(2317)^\pm)=2317\,\mathrm{MeV}$ (dashed).
The shaded bands indicate the one- and two-sigma regions of $f_-(0)/f_+(0)$ as measured by FOCUS \cite{Link:2004dh}.}
\label{RDplot}
\end{figure}

Substituting into Eq.~(\ref{GammaDtoK}), we obtain the values shown in Fig.~\ref{RDplot} for
$R_{D}^{-1} = \Gamma(D\rightarrow Ke\nu)/\Gamma(D\rightarrow K\mu\nu)$. 
Reading off from the graph, we assign the following $1\sigma$ and $2\sigma$ ranges
to $R_{D^0}$ and $R_{D^+}$:
\begin{equation}
(R_{D^0})^{-1} \;=\; (R_{D^+})^{-1} \;=\;
1.04\pm 0.02\;(1\sigma)\;,\quad 1.04^{+0.02}_{-0.06}\;(2\sigma)\;.
\end{equation}
%

\subsection{$D\rightarrow K^*\ell\nu$}

\begin{figure}[ht]
\includegraphics[width=8cm]{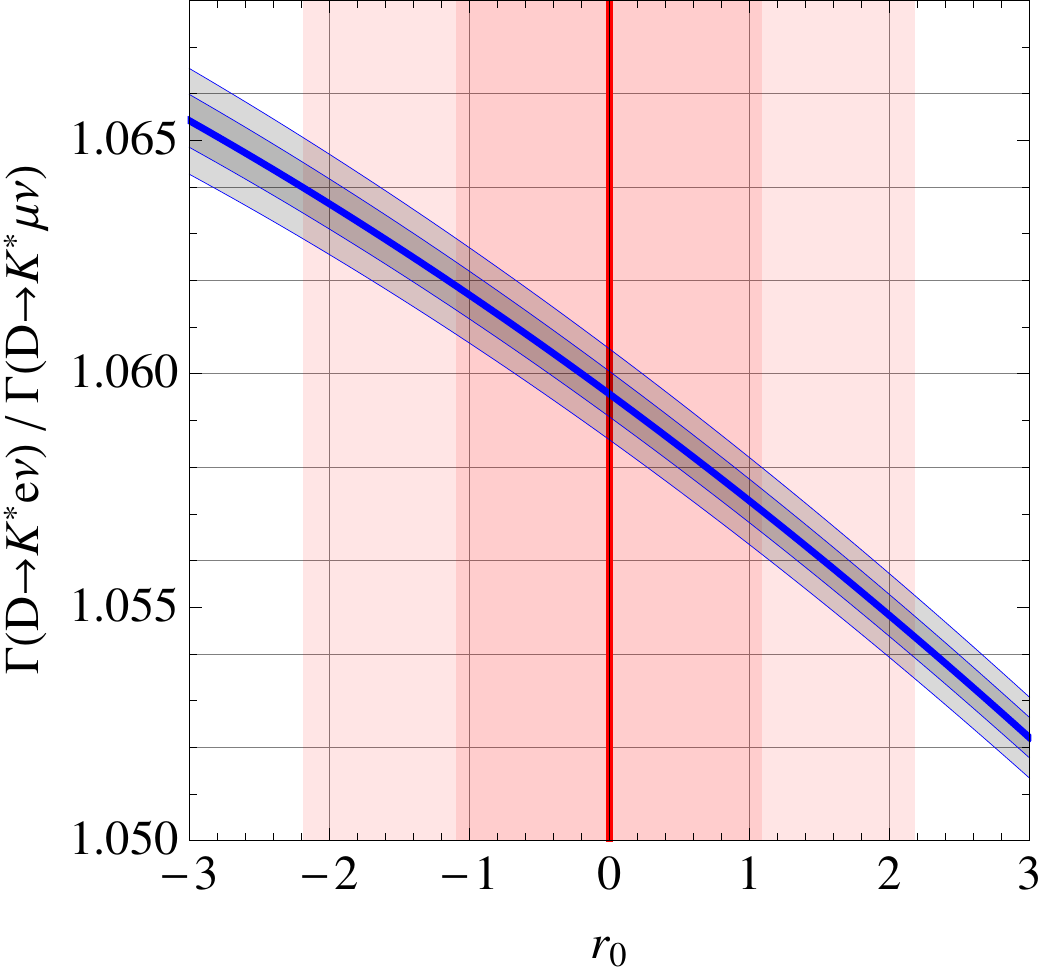}
\caption{The dependence of $(R^*_{D^+})^{-1}=\Gamma(D\rightarrow K^* e\nu)/\Gamma(D\rightarrow K^*\mu\nu)$ on 
the choice of $r_0$.  
The wide vertical bands are our assumed one- and two-sigma regions of $r_0$.
The narrow curved bands indicate the one and two-sigma uncertainties due to the error in $r_2$.
The uncertainty due to errors in $r_v$, $m_D$, and $m_{K^*}$ are negligible.}
\label{RDstarplot}
\end{figure}

Next, we consider the decay $D\rightarrow K^*\ell\nu_\ell$.
The form-factors are defined as \cite{DdecayFormFactors}
\begin{eqnarray}
\lefteqn{\bra{K^*(P_K,\epsilon)}\,\overline{s}\gamma_\alpha(1-\gamma_5)c\,\ket{D(P_D)}} \cr
& = & \varepsilon_{\alpha\beta\gamma\delta}P_D^\beta P_K^\gamma \epsilon^{*\delta}\;\dfrac{2}{m_D+m_{K^*}}\;V(q^2) \cr
& & +i\left\{
\epsilon_\alpha^*\left(m_D+m_{K^*}\right)\,A_1(q^2)
-\dfrac{(\epsilon^*\cdot q)}{m_D+m_{K^*}}(P_D+P_K)_\alpha \,A_2(q^2)
-\dfrac{(\epsilon^*\cdot q)}{q^2}\,q_\alpha (2m_{K^*})\, A_3(q^2) 
\right\}
\cr
& & +i\,\dfrac{(\epsilon^*\cdot q)}{q^2}\,q_\alpha (2m_{K^*})\, A_0(q^2)\;,
\end{eqnarray}
where
\begin{equation}
2m_{K^*} A_3(q^2) \;=\; (m_D+m_{K^*})\,A_1(q^2) - (m_D-m_{K^*})\,A_2(q^2)\;,\qquad
A_3(0)\;=\;A_0(0)\;,
\end{equation}
and $q^\alpha = (P_D-P_K)^\alpha$.
Here, $V(q^2)$ is the contribution of vector intermediate states, while $A_1(q^2)$ and $A_2(q^2)$ are
contributions of axial-vector intermediate states. $A_3(q^2)$ combines with $A_1(q^2)$ and $A_2(q^2)$
to maintain the transversality of these axial-vector contributions.  $A_0(q^2)$ is the contribution of
scalar intermediate states.  The condition $A_3(0)=A_0(0)$ is necessary for the cancellation of the poles
at $q^2=0$.
If we rewrite $A_3$ in terms of $A_1$ and $A_2$, the above expression becomes
\begin{eqnarray}
\lefteqn{\bra{K^*(P_K,\epsilon)}\,\overline{s}\gamma_\alpha(1-\gamma_5)c\,\ket{D(P_D)}} \cr
& = & \varepsilon_{\alpha\beta\gamma\delta}P_D^\beta P_K^\gamma \epsilon^{*\delta}\;\dfrac{2}{m_D+m_{K^*}}\;V(q^2) \cr
& & +i\left\{
\left(m_D+m_{K^*}\right)
\left[\,\epsilon_\alpha^* - \dfrac{(\epsilon^*\cdot q)\,q_\alpha}{q^2}\,\right]\,A_1(q^2)
-\dfrac{(\epsilon^*\cdot q)}{m_D+m_{K^*}}
\left[\,(P_D+P_K)_\alpha  - \dfrac{(m_D^2-m_{K^*}^2)\,q_\alpha}{q^2}\,\right]\,A_2(q^2)
\right\}
\cr
& & +i\,\dfrac{(\epsilon^*\cdot q)}{q^2}\,q_\alpha (2m_{K^*})\, A_0(q^2)\;,
\end{eqnarray}
making the transversality of the axial-vector terms manifest.
The decay width in terms of these form-factors is
\begin{eqnarray}
\Gamma & = &
\dfrac{G_F^2|V_{cs}|^2 m_D^3}{2^7\pi^3}
\int_{m_\ell^2}^{(m_D-m_{K^*})^2} dt \cr
& & \times
\Biggl[
\dfrac{4m_D^2}{3(m_D+m_{K^*})^2}\,
\dfrac{t}{m_D^2}\,
\dfrac{\lambda^{3/2}(m_D^2,m_{K^*}^2,t)}{m_D^6}
\left(1+\dfrac{m_\ell^2}{2t}\right) 
\left(1-\dfrac{m_\ell^2}{t}\right)^2
V(t)^2 
\cr
& & \quad +
\dfrac{2(m_D+m_{K^*})^2}{3 m_D^2}\,
\dfrac{t}{m_D^2}\,
\dfrac{\lambda^{1/2}(m_D^2,m_{K^*}^2,t)}{m_D^2}
\left[\;3+\dfrac{\lambda(m_D^2,m_{K^*}^2,t)}{4m_{K^*}^2 t}\,\right]
\left(1+\dfrac{m_\ell^2}{2t}\right) 
\left(1-\dfrac{m_\ell^2}{t}\right)^2
A_1(t)^2
\cr
& & \quad
-\dfrac{\left(m_D^2-m_{K^*}^2-t\right)}{3 m_{K^*}^2}\,
\dfrac{\lambda^{3/2}(m_D^2,m_{K^*}^2,t)}{m_D^6}
\left(1+\dfrac{m_\ell^2}{2t}\right) 
\left(1-\dfrac{m_\ell^2}{t}\right)^2
A_1(t) A_2(t) \;.
\cr
& & \quad +
\dfrac{m_D^4}{6(m_D+m_{K^*})^2 m_{K^*}^2}\;
\dfrac{\lambda^{5/2}(m_D^2,m_{K^*}^2,t)}{m_D^{10}}
\left(1+\dfrac{m_\ell^2}{2t}\right) 
\left(1-\dfrac{m_\ell^2}{t}\right)^2
A_2(t)^2 
\cr
& & \quad +
\dfrac{\lambda^{3/2}(m_D^2,m_{K^*}^2,t)}{m_D^6}
\;\dfrac{m_\ell^2}{t}
\left(1-\dfrac{m_\ell^2}{t}\right)^2
A_0(t)^2
\Biggr]\;,
\label{GammaDtoKstar}
\end{eqnarray}
where $t=q^2$.
The $A_0(t)$ contribution, being proportional to $m_\ell^2$, cannot be measured at current
experimental sensitivities. For the other form-factors, the following expressions motivated by
vector-meson-dominance \cite{Sakurai:1969ss} is assumed:
\begin{equation}
V(t)\;=\;V(0)\;\dfrac{1}{1-t/m_V^2}\;,\qquad
A_i(t)\;=\;A_i(0)\;\dfrac{1}{1-t/m_A^2}\;,\quad(i=1,2,3)\;.
\label{VMD2}
\end{equation}
The pole masses are taken to be 
$m_V = m(D^*_s{}^\pm) = 2.1\,\mathrm{GeV}$ and $m_A = m(D_{s1}(2460)^\pm) = 2.5\,\mathrm{GeV}$,
which are the masses of the lowest lying $D_s$-mesons with $J^P=1^-$ and $1^+$, respectively.
Experimental data is then fit with the ratios $r_v\equiv V(0)/A_1(0)$, and $r_2\equiv A_2(0)/A_1(0)$.
These ratios have been measured by various experiments via the decay sequence
$D^+\rightarrow \overline{K^*}(892)^0 \ell^+\nu_\ell$, $\overline{K^*}(892)^0\rightarrow K^-\pi^+$,
most recently by FOCUS \cite{Link:2002wg}, which also measured the decay sequence
$D^0\rightarrow K^*(892)^- \mu^+\nu_\mu$, $K^*(892)^-\rightarrow \overline{K^0}\pi^-$
\cite{Link:2004uk}.
The current world averages are \cite{Amsler:2008zzb}
\begin{eqnarray}
r_v & = & \dfrac{V(0)}{A_1(0)} \;=\; 1.62\pm 0.08\;, \cr
r_2 & = & \dfrac{A_2(0)}{A_1(0)} \;=\; 0.83\pm 0.05\;.
\end{eqnarray}
For our purpose, we also need to specify $A_0(t)$ which we assume is of the form
\begin{equation}
A_0(t) \;=\; A_3(0) + c_S\;\dfrac{t/m_S^2}{1-t/m_S^2}\;,
\end{equation}
with $m_S = m_{D_s^\pm} = 2.0\,\mathrm{GeV}$, the mass of the lowest lying $D_s$-meson with $J^P=0^-$.
Normalizing to $A_1(0)$, we have
\begin{equation}
\dfrac{A_0(t)}{A_1(0)}\;=\; r_3 + r_0\;\dfrac{t/m_S^2}{1-t/m_S^2}\;,
\label{VMD3}
\end{equation}
with
\begin{equation}
r_3 
\;=\; \dfrac{(m_D+m_{K^*})}{2m_{K^*}} - \dfrac{(m_D-m_{K^*})}{2m_{K^*}}\,r_2
\;=\; 1.09\pm 0.03\;.
\end{equation}
For $r_0$, we arbitrarily assume that its $1\sigma$ range is $r_0 = 0\pm r_3$.
Substituting Eqs.~(\ref{VMD2}) and (\ref{VMD3}) into Eq.~(\ref{GammaDtoKstar}), we obtain the
values shown in Fig.~\ref{RDstarplot} for the ratio
$(R^*_{D^+})^{-1}=\Gamma(D\rightarrow K^* e\nu)/\Gamma(D\rightarrow K^*\mu\nu)$.
Reading off from the graph, we assign the following 1$\sigma$ and $2\sigma$ ranges:
\begin{equation}
(R^*_{D^+})^{-1}\;=\;
1.060^{+0.002}_{-0.003}\;(1\sigma)\;,\qquad
1.060^{+0.005}_{-0.007}\;(2\sigma)\;.
\end{equation}


\end{document}